\RequirePackage{fix-cm}
\documentclass[natbib]{svjour3} % onecolumn (second format)
\smartqed % flush right qed marks, e.g. at end of proof
\usepackage{graphicx}
\usepackage{amsmath}
\usepackage{amssymb}
\usepackage{gensymb} % for \degree symbol
\usepackage[utf8]{inputenc} % for accentuated characters
\usepackage{rotating} % for rotated figures
\usepackage{caption} % for caption of side by side figures
\journalname{Celestial Mechanics and Dynamical Astronomy}
\begin{document}

\title{Long term dynamics beyond Neptune: secular models to study the regular motions}
%\titlerunning{Short form of title} % if too long for running head

\author{Melaine~Saillenfest$^{1,2}$ \and Marc~Fouchard$^{1,3}$ \and Giacomo~Tommei$^{2}$ \and Giovanni~B.~Valsecchi$^{4,5}$}
\authorrunning{Melaine Saillenfest \and Marc Fouchard \and Giacomo Tommei \and Giovanni B. Valsecchi} % if too long for running head

\institute{
   M. Saillenfest \at
   melaine.saillenfest@obspm.fr\\
%   \\
%   M. Fouchard \at
%   marc.fouchard@obspm.fr\\
%   \\
%   G. Tommei \at
%   giacomo.tommei@unipi.it\\
%   \\
%   G.B. Valsecchi \at
%   giovanni@iaps.inaf.it\\
   \\
   $^1$ IMCCE, Observatoire de Paris, 77 av. Denfert-Rochereau, 75014 Paris, France\newline
   $^2$ DM, Università di Pisa, Largo Bruno Pontecorvo 5, 56127 Pisa, Italia\newline
   $^3$ LAL-IMCCE, Université de Lille, 1 Impasse de l’Observatoire, 59000 Lille, France\newline
   $^4$ IAPS-INAF, via Fosso del Cavaliere 100, 00133 Roma, Italia\newline
   $^5$ IFAC-CNR, via Madonna del Piano 10, 50019 Sesto Fiorentino, Italia
}

\date{Received: 29 January 2016 / Revised: 15 April 2016 / Accepted: 10 May 2016}

\maketitle

\begin{abstract}
   Two semi-analytical one-degree-of-freedom secular models are presented for the motion of small bodies beyond Neptune. A special attention is given to trajectories entirely exterior to the planetary orbits. The first one is the well-known non-resonant model of~\cite{KOZAI_1962} adapted to the transneptunian region. Contrary to previous papers, the dynamics is fully characterized with respect to the fixed parameters. A maximum perihelion excursion possible of $16.4$ AU is determined. The second model handles the occurrence of a mean-motion resonance with one of the planets. In that case, the one-degree-of-freedom integrable approximation is obtained by postulating the adiabatic invariance, and is much more general and accurate than previous secular models found in the literature. It brings out in a plain way the possibility of perihelion oscillations with a very high amplitude. Such a model could thus be used in future studies to deeper explore that kind of motion. For complex resonant orbits (especially of type $1\!:\!k$), a segmented secular description is necessary since the trajectories are only ``integrable by parts". The two models are applied to the Solar System but the notations are kept general so that it could be used for any quasi-circular and coplanar planetary system.
   \keywords{secular model \and Lidov-Kozai mechanism \and mean-motion resonance \and transneptunian object}
\end{abstract}

\section{Introduction}
  The dynamical structure of the transneptunian region is still far from being fully understood, especially concerning high-perihelion objects and the link toward the Oort Cloud. For these objects, the orbital perturbations are very weak, both from inside (the planets) and from outside (passing stars and galactic tides). However, some of them are observed on very eccentric orbits (the most distant ones being Sedna and 2012VP$_\text{113}$ with eccentricities $0.85$ and $0.69$) which indicates that they have not formed in their present orbital state. Before thinking of exotic theories, an exhaustive survey has to be conducted on the different mechanisms that could produce such trajectories involving only what we take for granted about the Solar System dynamics, that is the orbital perturbations by the known planets and/or by galactic tides. The idea has been introduced in \cite{GALLARDO-etal_2012} and we will often refer to that article. For instance, it is known from a long time by numerical ways that the secular dynamics in a mean-motion resonance can produce high-amplitude oscillations of the perihelion distance \citep[see for example][]{GOMES-etal_2005}. Even if it is usually considered unlikely to produce objects of type Sedna \citep{MORBIDELLI-LEVISON_2004}, our goal is to characterize and quantify that kind of mechanism by other means than statistics on the output of numerical simulations.
  
  As a first step, we will focus in the present paper on planetary perturbations alone. The galactic tides, effective for very high semi-major axis \citep[see for instance][]{FOUCHARD-etal_2006}, are kept for future studies. We will further restrict the study on perihelion distances greater than the orbit of Neptune, that is on trajectories completely out of the planetary region. Such orbits can be divided into two broad classes:
  \begin{itemize}
     \item The first one, referred here as the Scattered Disc, contains the objects undergoing a diffusion of semi-major axis. It denotes a chaotic short time-scale dynamics, so these orbits are unstable by essence. It has be shown that a diffusive process is unable to produce a substantial variation in perihelion distance \citep[see][for a thorough review]{GALLARDO-etal_2012}. That kind of orbits will be dismissed in the present paper.
     \item The second class regroups the objects with integrable (or quasi-integrable) short time-scale dynamics. As such, their orbits can be described by secular models, which are nothing else than a first stage toward Arnold's action-angle variables (for a completely integrable motion). Such models can exhibit stable equilibrium points and libration zones for the secular argument of perihelion $\omega$ and perihelion distance $q$. If a particle follows such kind of orbit, we say that it experiences ``Lidov-Kozai mechanism", in reference to the pioneer papers of~\cite{KOZAI_1962,KOZAI_1985} and to the independent study of~\cite{LIDOV_1962} for the motion of artificial satellites. That class can be further divided into two kind of objects: the non-resonant ones (fixed semi-major axis) and those trapped in a mean-motion resonance with a planet (oscillating semi-major axis). To prevent any scattering, the non-resonant objects need a sufficiently high perihelion distance, the limit being estimated by~\cite{GALLARDO-etal_2012} as roughly $q_\text{min} = a/27.3 + 33.3$ AU (where $a$ stands for the semi-major axis expressed in AU). The resonant orbits are much more permissive because the forced link with one of the planets can act as a protective mechanism against diffusion. However, the resonance overlapping and, of course, the close encounters with Neptune, are still well-known sources of chaos for perihelion distances very close to the planetary region.
  \end{itemize}
  Note that these two broad classes are somehow permeable: a diffusion of semi-major axis can indeed stop abruptly with a resonance capture, or on the contrary, a quasi-integrable secular motion can get the perihelion distance to decrease toward the diffusion region. We will come back to that point later.
  
  The present work is devoted to the use of secular models to describe the trajectories of objects of the second class. That kind of models is widely used in celestial mechanics because it allows to study graphically in a glance a large variety of trajectories \citep[see for instance][]{MORBIDELLI_2002}. We give here a succinct context of its application in the region of interest: in 1962, Y. Kozai developed an analytical secular model for asteroids with arbitrary inclination and eccentricity. His model is designed for an external perturbing planet (namely Jupiter) and the article presents the dynamics given by the first terms of the analytical expansion. Then, \cite{KOZAI_1985} added the possibility of a mean-motion resonance between the particle and its perturber and turned to semi-analytical methods. As it assumes a fixed value of the resonant angle, that second model can only be used as a rough insight of the true resonant dynamics. Thanks to the increasing power of computers, \cite{THOMAS-MORBIDELLI_1996} used a semi-analytical approach to generalize the non-resonant model of Kozai for several planets. They presented a collection of secular level curves for semi-major axis greater than $30$ AU with a special attention given to perihelion distances inside the planetary region (the collision curves appear as pinch lines on the graphs). At last, \cite{GALLARDO-etal_2012} made a thorough review of the variety of trajectories beyond Neptune (see their introduction for a more detailed historical background). With a different approach, they obtained a Kozai-type analytical expansion for a set of internal perturbing planets (but only the very first terms are shown) and used it, as well as semi-analytical methods, to describe qualitatively the non-resonant dynamics for a perihelion outside the planetary region. They also modified the semi-analytical resonant model of \cite{KOZAI_1985} to deal with a more realistic sinusoidal evolution of the resonant angle, as already used in \cite{GOMES-etal_2005}. That method is however still unsatisfactory for a general study, since the evolution of mean-motion resonant angles in that region can actually undergo strong variations during the dynamics (centre, amplitude, frequency). These variations are besides unknown a priori. Some improvements have thus to be realised to take into account the precise variation of the resonant angle, in order to get an accurate representation of the dynamics rather than a rough approximation of it.
  
  To sum up, the background for secular dynamics beyond Neptune is now well established but the analysis found in the literature remain qualitative and incomplete. Since the quasi-integrable motion beyond Neptune is a promising mechanism to greatly modify the perihelion distance of small bodies, a special effort has to be deployed to construct secular models designed to explore in a straightforward and accurate way all the possible regular orbits. In this line of thinking, the aim of this work is twofold: provide a thorough analysis of the non-resonant case and develop an accurate resonant secular model\footnote{To prevent any confusion in the following, note that the present paper will never deal with the so-called ``secular resonances". What we call here a ``resonant secular model" is a secular model that takes into account a mean-motion resonance between the particle and one of the planets.}. The application to real objects and the possible implications concerning the distribution of the transneptunian orbits will be studied in future works.
  
  In Sec.~\ref{sec:model}, we present the planetary model used and the resulting osculating Hamiltonian function, starting point for any secular representation. In Sec.~\ref{sec:nres} we revisit Kozai's non-resonant secular model in the transneptunian region. Its general form is given for an arbitrary number of terms. An analysis of the lowest order terms, similar to the one of~\cite{GALLARDO-etal_2012}, is performed to get a time-scale information (typical duration of the Lidov-Kozai cycles). The semi-analytical method of~\cite{THOMAS-MORBIDELLI_1996} is then used to explore systematically the space of parameters for a perihelion beyond Neptune: all the behaviours allowed by a non-resonant secular dynamics are thus described and quantified in an exhaustive way. Then, Section~\ref{sec:res} presents the construction of the resonant secular model. It is explained why previous models, which assume a particular evolution of the resonant angle \citep{KOZAI_1985,GALLARDO-etal_2012,BRASIL-etal_2014}, can be inaccurate or cumbersome. In the present paper, the adiabatic invariant theory is used to get a one-degree-of-freedom secular system. That strategy turns out to be effective to study the resonant dynamics beyond Neptune and as aesthetic as non-resonant models: all the possible orbits are described by the level curves of a secular Hamiltonian with two free parameters. Finally, Section~\ref{sec:ex} shows some illustrations of the resonant model, along with detailed explanations about its use for the various types of dynamics we can be confronted with. As the variety of trajectories is found to be much more complex and richer than in the non-resonant case, an exhaustive exploration of the parameter space is left for future works.
  
\section{Planetary model and Hamiltonian function}\label{sec:model}
   In a heliocentric reference frame, the Hamiltonian function for a test-particle undergoing the gravitational attraction of the Sun and $N$ planets writes:
   \begin{equation}\label{eq:Hfirst}
      \mathcal{H}(L,G,H,l,g,h,t) = -\frac{\mu^2}{2L^2} - \sum_{i=1}^{N}\mu_i\left(\frac{1}{|\mathbf{r}-\mathbf{r}_i|} - \mathbf{r}\cdot\frac{\mathbf{r}_i}{|\mathbf{r}_i|^3}\right)
   \end{equation}
   where $\mathbf{r}$ and $\mathbf{r}_i$ are the heliocentric positions of the particle and of the $i^\text{th}$ planet, and $\mu$ and $\mu_i$ are the gravitational constant times the masses of the Sun and the $i^\text{th}$ planet, respectively. Written in that form, $\mathcal{H}$ is time-dependent through the planetary positions, supposed known functions of the time: $\mathbf{r}_i \equiv \mathbf{r}_i(t)$. The Hamiltonian function~\eqref{eq:Hfirst} is written in Delaunay canonical coordinates:
   \begin{equation}\label{eq:Delcoo}
      \left\{
      \begin{aligned}
         l &= M \\
         g &= \omega \\
         h &= \Omega
      \end{aligned}
      \right.
      \text{\ \, \ and\ \ }
      \left\{
      \begin{aligned}
         L &= \sqrt{\mu a}\\
         G &= \sqrt{\mu a\,(1-e^2)}\\
         H &= \sqrt{\mu a\,(1-e^2)}\,\cos I
      \end{aligned}
      \right.
   \end{equation}
   where $\{a,e,I,\omega,\Omega,M\}$ are the usual heliocentric keplerian elements, related to $\mathbf{r}$ via:
   \begin{equation}
      \mathbf{r} =
      r
      \begin{pmatrix}
         \cos(\omega+\nu)\cos\Omega-\sin(\omega+\nu)\sin\Omega\cos I \\
         \cos(\omega+\nu)\sin\Omega+\sin(\omega+\nu)\cos\Omega\cos I \\
         \sin(\omega+\nu)\sin I
      \end{pmatrix}
   \end{equation}
   with:
   \begin{equation}
      |\mathbf{r}| = r \equiv \frac{a\,(1-e^2)}{1+e\cos\nu}
      \text{\ \ \ and\ \ \ }\nu\equiv \nu(e,M)\text{\ \ \ from Kepler equation.}
   \end{equation}
   As usual, we split $\mathcal{H}$ into its keplerian part $\mathcal{H}_0$ and the planetary perturbations $\varepsilon\,\mathcal{H}_1$, where the size $\varepsilon$ of the perturbation is related to $\max\,\{m_i\}=m_\text{J}$:
   \begin{equation}
      \mathcal{H}(L,G,H,l,g,h,t) = \mathcal{H}_0(L) + \varepsilon\,\mathcal{H}_1(L,G,H,l,g,h,t)
   \end{equation}
   In order to study the specific role of each planet, we must now choose a planetary model, that is an explicit formulation of the $\{\mathbf{r}_i(t)\}$ functions. This can be done either by a synthetic representation or by analytical expansions as in~\cite{LEMAITRE-MORBIDELLI_1994} or~\cite{MOONS-etal_1998}. We will opt here for the very simple planetary model used by~\cite{KOZAI_1962} and many others thereafter, where the $N$ planets evolve on circular and coplanar orbits. As recalled by~\cite{THOMAS-MORBIDELLI_1996}, such a model can be seen as the dominant term of an expansion in powers of the planetary eccentricities and inclinations. Anyway, that approximation seems quite viable, given that the only relevant planetary perturbations in the region under study come from the four giant planets (eccentricities $<0.1$ and inclinations $<3$\degree), on relatively stable orbits from the end of the planetary migration \citep[see for instance][]{LASKAR_1988,LASKAR_1990,TSIGANIS_2005}.
   
   With that planetary model, it is straightforward to disentangle the effect of each planet, since:
   \begin{equation}\label{eq:pmod}
      \mathbf{r}_i(t) = a_i
      \begin{pmatrix}
         \cos\lambda_i(t) \\
         \sin\lambda_i(t) \\
         0
      \end{pmatrix}
      \text{\ \ \ with\ \ \ }
      \lambda_i(t) = n_i\,t+\lambda_{i0}
   \end{equation}
   where the semi-major axis $a_i$ is constant and $n_i^2\,a_i^3 = \mu + \mu_i$. We then get rid of the explicit time dependence by defining the angles $\{\lambda_i\}$ as new canonical coordinates, along with their conjugated momenta $\{\Lambda_i\}$ artificially added to the non perturbed part $\mathcal{H}_0$:
   \begin{equation}\label{eq:H0}
      \mathcal{H}_0 = -\frac{\mu^2}{2L^2} + \sum_{i=1}^{N}n_i\,\Lambda_i
   \end{equation}
   The general form of the Hamiltonian function writes finally:
   \begin{equation}\label{eq:Hgen}
      \mathcal{H}\Big(\{\Lambda_i\},L,G,H,\{\lambda_i\},l,g,h\Big) = \mathcal{H}_0\Big(\{\Lambda_i\},L\Big) + \varepsilon\,\mathcal{H}_1\Big(L,G,H,\{\lambda_i\},l,g,h\Big)
   \end{equation}
   It is the starting point for all the models presented below (non-resonant and resonant ones).

\section{Non-resonant case}\label{sec:nres}
   In order to switch to secular coordinates and compute the secular Hamiltonian function, a choice has now to be made: with $\mathcal{H}$ as described above, it is indeed impossible to get rid of the short periodic terms analytically without the use of infinite series. Section~\ref{sec:nres} is thus organised as follows: in Sec.~\ref{subsec:anmod}, the analytical model of~\cite{KOZAI_1962} is adapted to the outer Solar System (several interior planets). The dominant terms are then studied in Sec.~\ref{subsec:fta}. Naturally, this will give only a rough picture of the secular dynamics, but some general results will be obtained and guide the construction of an ``exact" semi-analytical model in Sec.~\ref{subsec:sanmod}. That last model is not new and it has already been applied to transneptunian objects \citep[see for instance][]{THOMAS-MORBIDELLI_1996,GALLARDO-etal_2012}. Some aspects are still worth to be detailed, however, to depict a general picture of the non-resonant dynamics for a perihelion outside the orbit of Neptune and compare it later to the resonant case.

   \subsection{Analytical model}\label{subsec:anmod}
   Let us recall here the method of~\cite{KOZAI_1962}. The possible large eccentricities and inclinations of the transneptunian objects make impossible the use of a classical expansion around a circular orbit in the planetary plane (we are indeed outside or very close to its radius of convergence). A development centred on some specific values \citep[see for instance][]{ROIG-etal_1998} is also inappropriate because of the possible large variations of orbital elements, and because it would imply a loss of generality. However, considering that all the semi-major axis concerned are greater than Neptune's, we can think of a development in Legendre polynomials, that is in powers of the $\{r_i/r\}$ ratios\footnote{The expansion of~\cite{KOZAI_1962} makes use of the inverse ratio: he was indeed interested of trajectories entirely \emph{interior} to the orbit of Jupiter.}:
   \begin{equation}\label{eq:devPL}
      \frac{1}{|\mathbf{r}-\mathbf{r}_i|} = \frac{1}{r}\sum_{n=0}^{+\infty}\left(\frac{r_i}{r}\right)^n P_n(\cos\psi_i)
   \end{equation}
   where $P_n(x)$ is the $n^\text{th}$ Legendre polynomial, and the angle $\psi_i$ is defined by :
   \begin{equation}
      \cos\psi_i = \widehat{(\mathbf{r},\mathbf{r}_i)} = \frac{\mathbf{r}\cdot\mathbf{r}_i}{r\,r_i}
   \end{equation}
   Naturally, that kind of development restricts us to trajectories that never come inside the planetary region, that is, for perihelion distances $q=a(1-e)$ always greater than $\max\,\{a_i\}=a_\text{N}$. This is not of great concern as we are precisely looking for orbits entirely exterior to the planets, but we must keep in mind that the convergence will be very poor for a perihelion near the orbit of Neptune.
   
   Let us now switch to secular coordinates. To do so, we will use the classical Lie-series formalism for the perturbed Hamiltonian system~\eqref{eq:Hgen}: since we assume in that section that there is no mean-motion resonance in the system, the fast angles $l$ and $\{\lambda_i\}$ can be removed by a close-to-identity transformation. In the secular coordinates, the Hamiltonian function writes then:
   \begin{equation}
      \mathcal{F} = \mathcal{F}_0 + \varepsilon\,\mathcal{F}_1 + \mathcal{O}(\varepsilon^2)
   \end{equation}
   where $\mathcal{F}_0$ is \emph{functionally} equal to $\mathcal{H}_0$ and $\varepsilon\,\mathcal{F}_1$ is \emph{functionally} equal to the average of $\varepsilon\,\mathcal{H}_1$ with respect to the independent angles $l$ and $\lambda_1,\lambda_2...\lambda_N$. In the region considered here, we judge enough to carry on the transformation up to the first order of $\varepsilon$. Note that we will actually never compute the change of coordinates, but simply suppose its existence.
   
   The indirect part of the perturbation vanishes under the average over $\lambda_i$, and the double integration of the direct part can be computed analytically thanks to the simple planetary model~\eqref{eq:pmod} and the development~\eqref{eq:devPL}:
   \begin{equation}\label{eq:eF1}
      \varepsilon\,\mathcal{F}_1 = -\sum_{i=1}^{N}\frac{1}{4\pi^2}\int_{0}^{2\pi}\!\!\!\int_{0}^{2\pi}\!\!\!\frac{\mu_i}{|\mathbf{r}-\mathbf{r}_i|}\,\mathrm{d}\lambda_i\,\mathrm{d}l = -\frac{1}{a}\sum_{n=0}^{+\infty}\left(\sum_{i=1}^{N}\mu_i\left(\frac{a_i}{a}\right)^{2n}\right)\mathcal{B}_{n}(e,I,\omega)
   \end{equation}
   where $\mathcal{B}_0 = 1$, and for $n>0$ the $\mathcal{B}_n$ functions are of the form:
   \begin{equation}\label{eq:Bn}
      \mathcal{B}_n(e,I,\omega) = \frac{\alpha_n}{(1-e^2)^\frac{4n-1}{2}}\,\sum_{k=0}^{n-1}\,P_n^k(e)\times Q_n^k(\cos I)\times \,e^{2k}\sin^{2k}\!\!I\,\cos(2k\omega)
   \end{equation}
   In that expression, $\alpha_n$ is a rational coefficient and $P_n^k$ and $Q_n^k$ are even polynomials of order $2(n-k-1)$ and $2(n-k)$ respectively. The explicit expressions of the first terms are given in the appendix, as well as some computation details. Note that the variables $(a,e,I,\omega)$ should then be replaced by their expressions in Delaunay elements~\eqref{eq:Delcoo} to get the Hamiltonian in canonical coordinates. Its general expression is thus (at first order of the planetary masses):
   \begin{equation}\label{eq:Fgen}
      \mathcal{F}\Big(\{\Lambda_i\},L,G,H,g\Big) = \mathcal{F}_0\Big(\{\Lambda_i\},L\Big) + \varepsilon\,\mathcal{F}_1\Big(L,G,H,g\Big)
   \end{equation}
   where $\mathcal{F}_0$ is given by~\eqref{eq:H0} and $\varepsilon\,\mathcal{F}_1$ by~\eqref{eq:eF1}. Please note that even if we write the coordinates with the same symbols as before, we now manipulate the \emph{secular} coordinates, related to the osculating ones by a complex canonical transformation.
   
   First of all, one can see that the angle $h=\Omega$ has disappeared during the transformation. This happened because of the symmetry of rotation implied by the circular and coplanar planetary orbits. Furthermore, the secular Hamiltonian depends only on the magnitude of $H/G=\cos I$ (not its sign), and it is $\pi$-periodic and symmetric with respect to $\pi/2$ in $g=\omega$.
   
   The analysis of the non-resonant secular dynamics is rather simple because we are left with only one degree of freedom: the secular momenta $L$ and $\Lambda_1,\Lambda_2...\Lambda_N$ are conserved, as well as $H$ thanks to the extra disappearance of $h$. So, all the possible orbits can be described by plotting the level curves of $\mathcal{F}$ in the $(G,g)$ plane with $L$ and $H$ as free parameters (the $\mathcal{F}_0$ part is constant and can be omitted). For a more direct interpretation, the secular Hamiltonian $\mathcal{F}$ will actually be drawn in the $(q,\omega)$ plane, equivalent to the $(G,g)$ plane. The two parameters will also be rewritten as:
   \begin{equation}
      \left\{
      \begin{aligned}
         a &= L^2/\mu \\
         C_K &= (H/L)^2 = (1-e^2)\cos^2I
      \end{aligned}
      \right.
   \end{equation}
   where we call $C_K$ the ``Kozai constant". Note that we chose to square the $H/L$ ratio to stress the independence of $\mathcal{F}$ over its sign. That constant links the secular eccentricity and inclination of the particle, bound to comply with its level curves. The variations allowed by the value of $C_K$ are then:
   \begin{equation}
      e\in\Big[0\,,\sqrt{1-C_K}\Big]\text{\ \ \ and\ \ \ } \cos^2I\in\Big[C_K\,,1\Big]
   \end{equation}
   In order to explore the phase space with respect to the two parameters, let us remark at first that for a circular orbit, the secular Hamiltonian becomes also independent of $g=\omega$. The elements $(a,e,I)$ are thus constant, and the angles $\Omega$ and $\omega$ (ill-defined in that case) circulate.
   
   \subsection{Analysis of the lowest order terms}\label{subsec:fta}
   For more interesting orbits ($e\neq 0$), we will now look for possible equilibrium points. A rough insight of the non-resonant secular dynamics can be obtained by a truncation of the development~\eqref{eq:eF1}. Such a simplified model was used for instance by~\cite{KINOSHITA-NAKAI_2007} to work out an analytical solution of Kozai's original problem (a single exterior perturber). In their case, the first terms are somehow simpler because a Legendre development for the inverse ratio does not imply the eccentricity in denominator as in~\eqref{eq:Bn}. Their small parameter contains besides a single planet. Naturally, some coefficients are similar, though, as they come directly from the Legendre polynomials.
   
   In our case, the general form of Eqs.~(\ref{eq:eF1},\ref{eq:Bn}) makes obvious that the truncated model will be accurate only for high semi-major axis and small eccentricities (that is, for trajectories always far from Neptune). Dropping the constant parts
   and carrying the expansion up to the very first term containing the angle $g$, we get
   (see appendix):
   \begin{equation}\label{eq:Ftrunc}
      \begin{aligned}
         \mathcal{F}(G,g) =\ &\delta^2\,\frac{1}{8}\,\left(\frac{L}{G}\right)^3\left(1-3\left(\frac{H}{G}\right)^2\right) \\
         +\ &\delta^4\,\frac{9}{1024}\,\left(\frac{L}{G}\right)^7\left[\Big(-3+30\left(\frac{H}{G}\right)^2-35\left(\frac{H}{G}\right)^4\Big)\Big(5-3\left(\frac{G}{L}\right)^2\Big) \right.\\
         &\left.\hspace{1.8cm}+\,10\,\Big(1-7\left(\frac{H}{G}\right)^2\Big)\Big(1-\left(\frac{H}{G}\right)^2\Big)\Big(1-\left(\frac{G}{L}\right)^2\Big)\cos(2g)\right] \\
         +\ &\mathcal{O}(\delta^6)
      \end{aligned}
   \end{equation}
   where we wrote symbolically:
   \begin{equation}
      \delta^{2n} \equiv \frac{1}{a}\sum_{i=1}^{N}\mu_i\left(\frac{a_i}{a}\right)^{2n}
   \end{equation}
   One can notice that the angle $g$ appears an order higher than in Kozai's original problem. From the Hamiltonian~\eqref{eq:Ftrunc}, the condition of stationarity writes:
   \begin{equation}\label{eq:stat}
      \left\{
      \begin{aligned}
         &\dot{g} = 0 + \mathcal{O}(\delta^4) \\
         &\dot{G} = 0 + \mathcal{O}(\delta^6)
      \end{aligned}
      \right.
      \iff
      \left\{
      \begin{aligned}
         &\cos^2\!I = 1/5 \\
         &\sin(2\omega) = 0
      \end{aligned}
      \right.
   \end{equation}
   The equilibrium points correspond thus to two very specific values of the inclination (about $63.4\degree$ or $116.6\degree$) and of the argument of perihelion ($0$ or $\pi/2\mod{\pi}$). These inclinations can be reached only for a parameter $C_K \leqslant 1/5$ (whereas the analogous limit in Kozai's original problem is $3/5$). The stability of the equilibrium points is given by the eigenvalues of the Jacobian matrix: we show easily that $g=0$ is a saddle point and $g=\pi/2$ is a central point. Finally, the imaginary parts of the eigenvalues give the frequency for small oscillations around the stable equilibrium:
   \begin{equation}
      \nu_{\pm} = \pm\frac{3}{1000}\sqrt{\frac{3}{5}}\frac{L^4}{H^6}\sqrt{\delta^2\,\delta^4\,(L^2-5H^2)} + \mathcal{O}(\delta^4)
   \end{equation}
   with $(L^2-5H^2)\geqslant 0$ because of~\eqref{eq:stat}.
   
\begin{figure}
      \centering
      \includegraphics[width=0.7\textwidth]{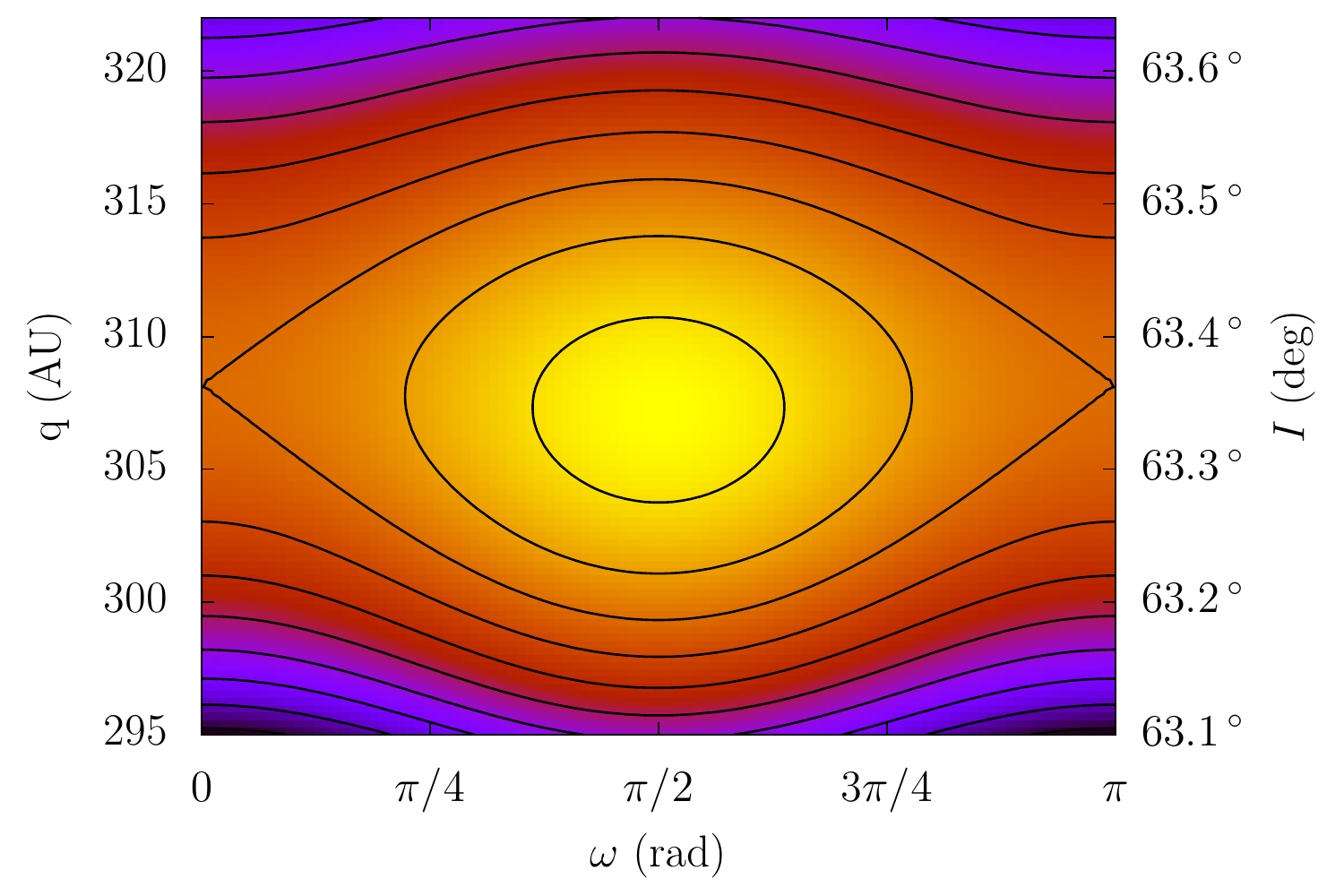}
      \caption{Level curves of the truncated version of $\mathcal{F}$ with terms up to $\delta^4$ (parameters : $a = 400$ AU, $C_K = 0.19$). The inclinations on the right are deduced from $q$ by $a$ and $C_K$ and are equivalent to $(116.9\degree,116.8\degree...116.4\degree)$, from bottom to top.}
      \label{fig:truncF}
   \end{figure}

\begin{figure}
      \includegraphics[width=\textwidth]{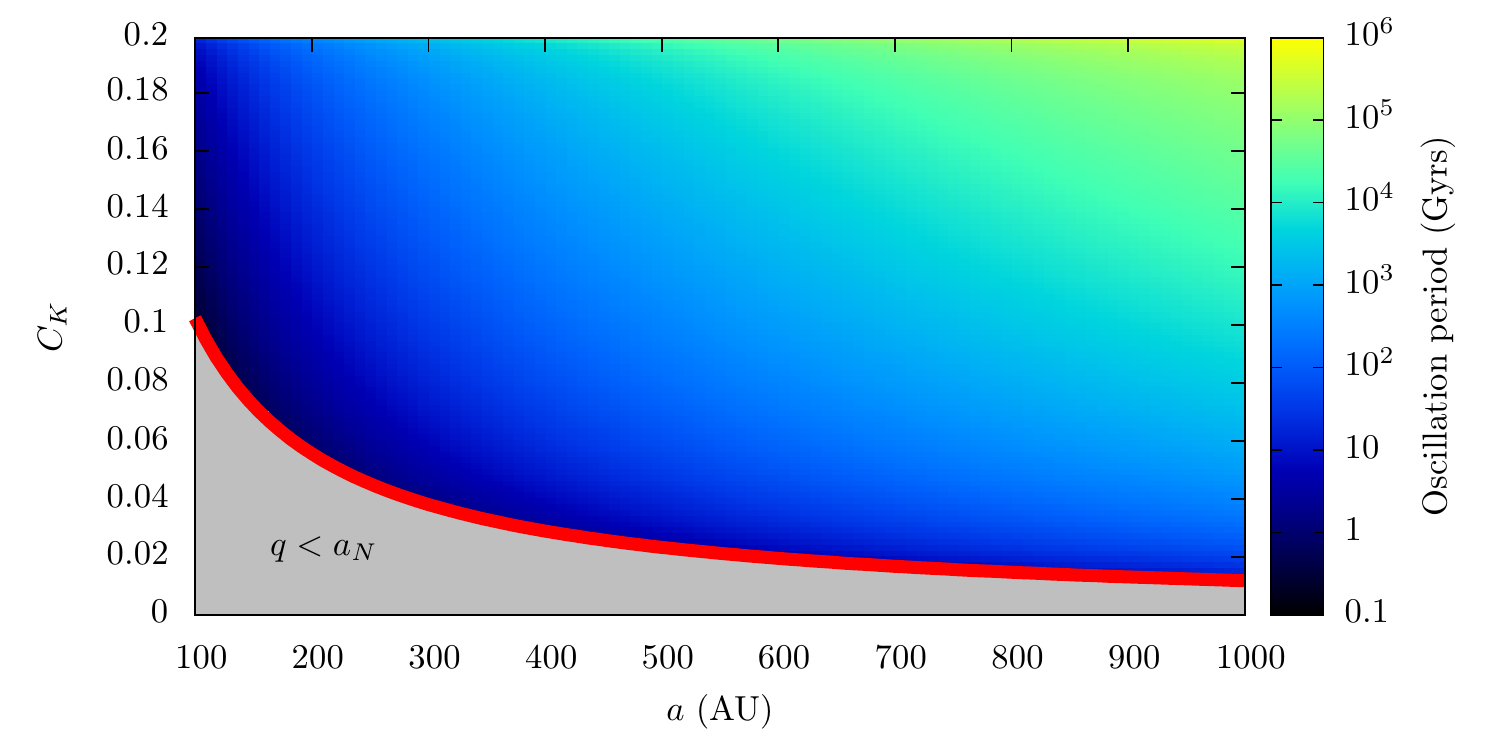}
      \caption{Oscillation period for small oscillations around the stable equilibrium. The red line defines the limit of convergence of the Legendre development (that is $q=a_\text{N}$).}
      \label{fig:nu}
   \end{figure}

   Figure~\ref{fig:truncF} gives an example of level curves obtained from the truncated secular Hamiltonian~\eqref{eq:Ftrunc}. The equilibrium is not located exactly at $I=63.4\degree$ because we neglected the term of order $\delta^4$ for $\dot{g}$ in Eq.~\eqref{eq:stat}: taking that term into account (or considering the infinite series as in Sec.~\ref{subsec:sanmod}), the inclination at equilibrium is actually a function of $a$ and $C_K$. As for the following, the model is here applied to Jupiter, Saturn, Uranus and Neptune ($N=4$), the mass of the inner planets being added to the Sun. Figure~\ref{fig:nu} shows the period of oscillation around the stable equilibrium as a function of the two parameters. On the red line, the perihelion at equilibrium is equal to Neptune's semi-major axis. Then, it goes up with $C_K$, until it reaches $a$ for $C_K=1/5$. We remark that the secular time-scale in that region is almost always greater than a billion years, which prevents probably any occurrence of a secular resonance with the planets. This is a new argument in support of a very simple planetary model (with fixed orbital elements) and is consistent with the results of~\cite{KNEZEVIC-etal_1991}.
   
   \subsection{Generalisation : semi-analytical model}\label{subsec:sanmod}
   In the previous part, we saw that it is possible to construct an analytical development of the non-resonant secular Hamiltonian function in powers of the $(r_i/r)$ ratios. The analysis of the first terms, then, led to rough general results about the geometry of the phase space. Naturally, these results are asymptotic (accurate only for high semi-major axis and small eccentricities), and cannot be used for trajectories near or inside the planetary region. In particular, it is known since~\cite{GALLARDO-etal_2012} that the oscillation island at $\omega=\pi/2$ disappears below some value of the semi-major axis, and that the equilibrium at $\omega=0$ can become stable.
   
   In order to get quantitative and accurate results, one can turn to numerical methods to compute the double average of $\varepsilon\,\mathcal{H}_1$: we thus get its exact value, that is the value obtained for an infinite number of terms in the Legendre development. In that section and the rest of the paper, we will use the integration package of~\cite{PIESSENS-etal_1983}, already successfully applied to such problems by~\cite{THOMAS-MORBIDELLI_1996} and~\cite{GRONCHI-MILANI_1999}. Each evaluation of $\mathcal{F}$ on a point $(\omega,q)$ will now require the numerical evaluation of the double integral~\eqref{eq:eF1}. Please note, however, that the general features of the secular Hamiltonian still hold (Eq.~\ref{eq:Fgen} and comments thereafter), and will help us to apprehend the geometry of the phase space.
   
\begin{figure}
      \centering
      \includegraphics[width=0.8\textwidth]{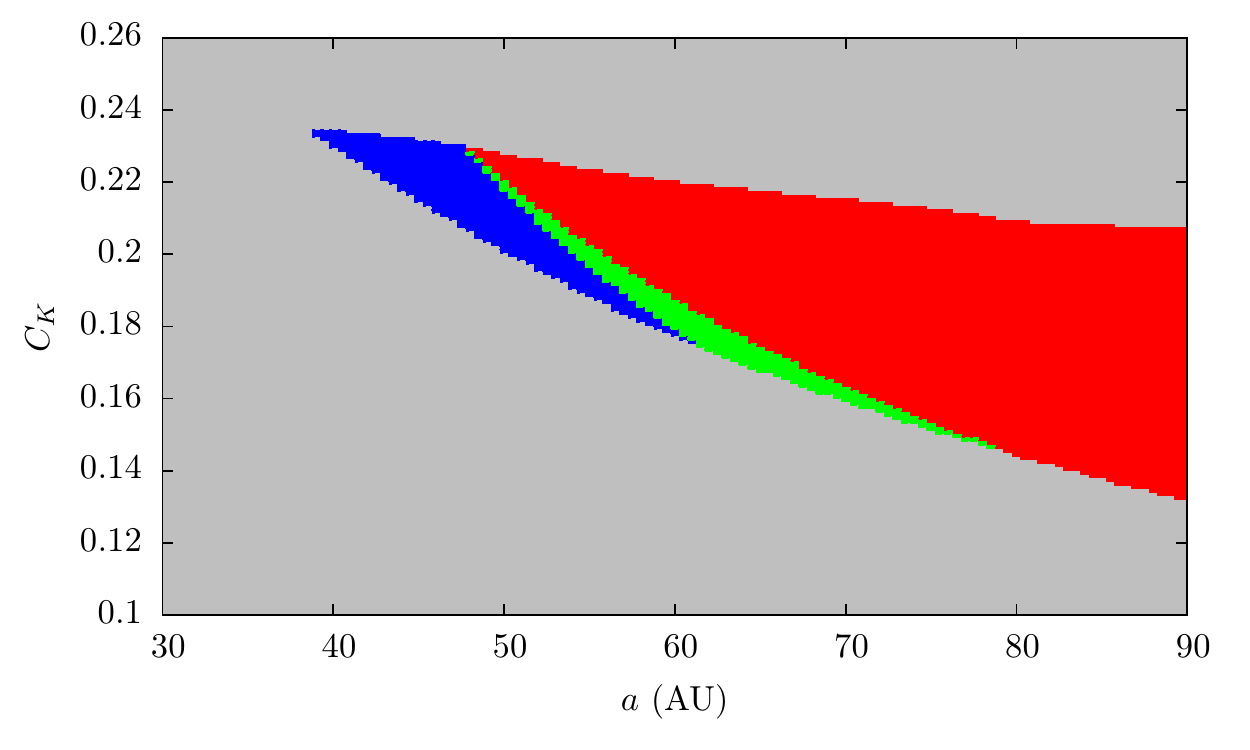}
      \caption{General geometry of the phase space with respect to the two parameters. The grey region denotes the absence of any equilibrium point for a perihelion greater than Neptune's semi-major axis. The blue region stands for the presence of a stable equilibrium point at $\omega=0$, the red one for a stable equilibrium point at $\omega=\pi/2$, and the green region for the simultaneous existence of both. For higher semi-major axis, the red region fills progressively the graph (see Fig.~\ref{fig:width} for a wider scale).}
      \label{fig:geometry}
   \end{figure}

   At this point, it seems vain to overcharge the article with new plots of the non-resonant secular regime beyond Neptune, since it is relatively well known from the work of~\cite{GALLARDO-etal_2012}. In that part, we will thus present only general results about the non-resonant dynamics by a systematic exploration of the parameter space\footnote{Even if the semi-analytical model is also valid for a perihelion inside the planetary region, we still limit the study to $q>a_\text{N}$ as this is the region of interest in the scope of this paper. For details about the non-resonant secular dynamics with a perihelion \emph{inside} the planetary region, see~\cite{THOMAS-MORBIDELLI_1996} or~\cite{GALLARDO-etal_2012}.}. Figure~\ref{fig:geometry} shows that the first terms analysis remains \emph{qualitatively} relevant for a semi-major axis greater than about $80$ AU: the equilibrium point at $\omega=\pi/2$ is indeed the only one to remain stable. In other words, the phase space is filled with circulation zones of $\omega$, where the perihelion oscillates with a very small amplitude. The only substantial variations of $q$ are located near that stable equilibrium, where $\omega$ can oscillate (see Sec.~\ref{subsec:fta}).
   
\begin{figure}
      \centering
      \includegraphics[width=\textwidth]{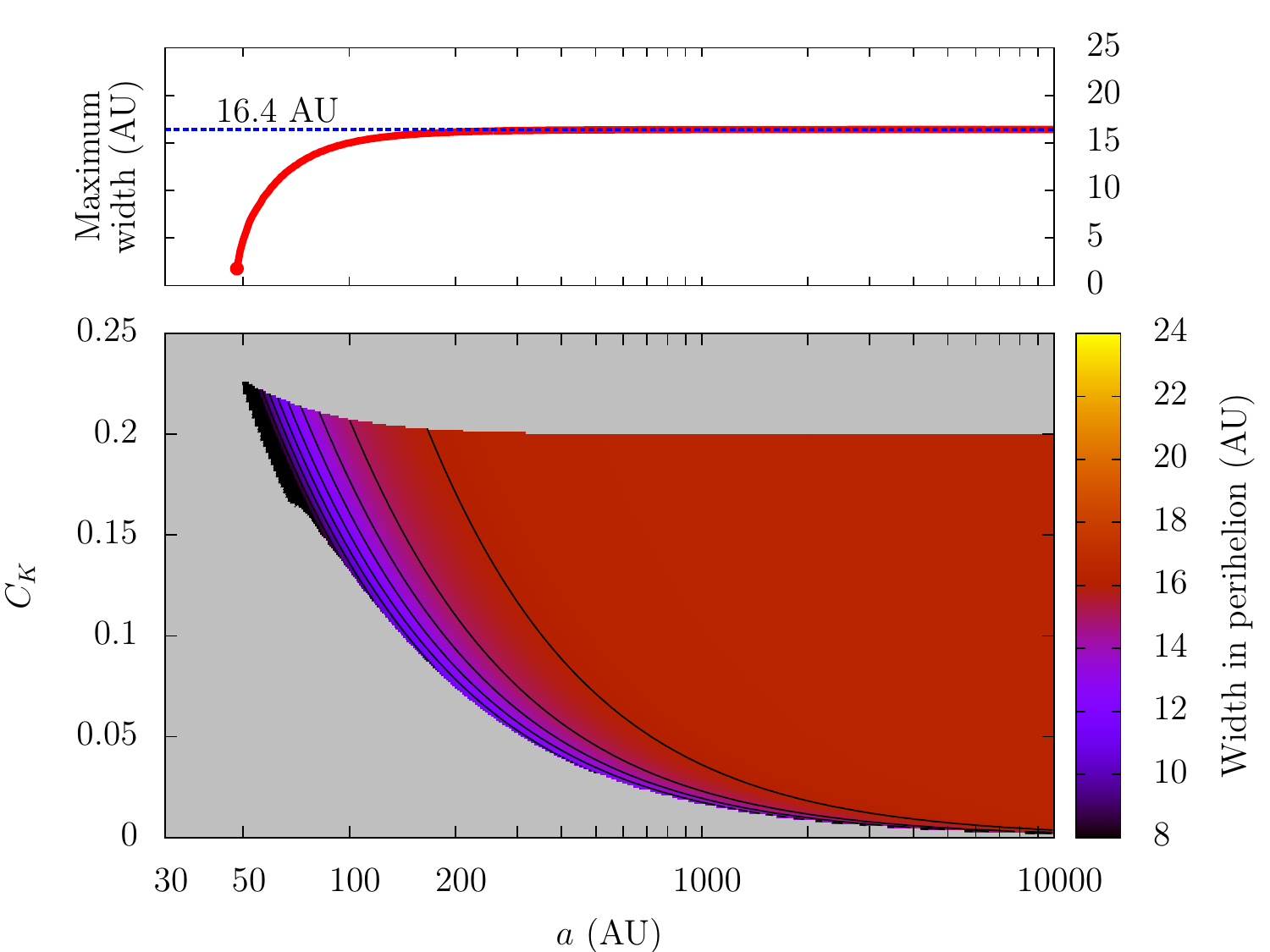}
      \caption{Width of the oscillation island around the stable equilibrium point at $\omega=\pi/2$. On the top graph, only the maximum value for all $C_K$ is retained. The grey region denotes the absence of such equilibrium point for a perihelion greater than Neptune's semi-major axis (or regions where the equilibrium point is so close to it that the lower separatrix ends below). The black lines are iso-width curves, plotted for every integer value (the upper one corresponds to $16$ AU). There is an asymptotic value of $q\approx 16.4$ AU, filling progressively all the graph when $a$ increases (the colour shade stops on red). The bump around $a=70$ AU marks the disappearance of the $\omega=0$ equilibrium point (see Fig.~\ref{fig:geometry}).}
      \label{fig:width}
   \end{figure}

   In order to define ``how substantial" it is, we used the semi-analytical approach to determine the exact width of the island with respect to the two parameters. The result is shown on Fig.~\ref{fig:width}: for each value of the parameters $(a,C_K)$, we searched numerically for the position of the saddle point, and then followed the two separatrices until they reached their maximum deployment. In the grey areas, there is no equilibrium point possible for a perihelion beyond Neptune's orbit: in particular, we notice that the upper limit of $C_K=1/5$ obtained analytically is rather well respected (and almost exact for $a>300$ AU). Moreover, the inclination at equilibrium was never found to be distant by more than $3\degree$ from the rough analytical value of Sec.~\ref{subsec:fta}. Then, the important point of Fig.~\ref{fig:width} is the existence of an asymptotic maximum width of the oscillation island of about $16.4$ AU. Since this result is only numerical, there is actually no way at this point to determine if it is a true asymptote or if the rate of increase tends just to a very small value\footnote{Note that an analytical search for the two separatrices at $\omega=\pi/2$ using an expansion of~\eqref{eq:Ftrunc} at order 2 of $G$ around the equilibrium does show an asymptotic flat width at about $16.4065975$ AU.}. However, this is not of great concern because a semi-major axis greater than some tens of thousands AU looses obviously its physical meaning (notice the log-scale on Fig.~\ref{fig:width}). Thus, if a particle begins with an initial perihelion near Neptune (say $35$ AU), the very maximum value it could reach in the future with that mechanism would be of about $50$ AU. The excursion is consequent but still well below the perihelion distances of Sedna and 2012VP$_{113}$. Furthermore, we saw in Sec.~\ref{subsec:fta} that the oscillation island is very narrow in terms of inclination (near $63.4\degree$ or $116.6\degree$) which restricts severely the probability for an object to undergo that kind of process.
   
\section{Resonant case}\label{sec:res}
   If the particle presents a mean-motion resonance with one of the planets, the coordinate change used in Sec.~\ref{sec:nres} to get the secular coordinates is not defined any more (some terms explode in the neglected part of the Lie series). A particular treatment for the resonant terms is thus required. Let us consider a single resonance of type $k_p\!:\!k$ with a resonant angle of the form:
   \begin{equation}\label{eq:sigma}
      \sigma = k\,\lambda - k_p\,\lambda_p - (k-k_p)\,\varpi
      \text{\ \ \ ,\ \ \ } k,k_p\in\mathbb{N}\text{\ \ \ ,\ \ \ }k>k_p
   \end{equation}
   In this expression, the angles $\lambda$ and $\lambda_p$ refer to the mean longitudes of the particle and of the planet $p$ involved, and $\varpi=\omega+\Omega$. Because the planets are supposed on circular and coplanar orbits, no other planetary angle can appear. Concerning the other possible angles associated with the $k_p\!:\!k$ resonance (those involving a further term in $\Omega$), they can be studied just as we will show for the angle $\sigma$: the method is quite general and can be applied to a large variety of dynamical systems. The only feature we need in order to define a suitable secular Hamiltonian is a clear hierarchy between the time-scales. In our case, we have now three of them:
   \begin{itemize}
      \item[$\bullet$] the short periods ($M$ and $\lambda_1,\lambda_2...\lambda_N$)
      \item[$\bullet$] the semi-secular periods (oscillation of the resonant angle $\sigma$)
      \item[$\bullet$] the secular periods (precession of $\omega$ and $\Omega$), that is the Lidov-Kozai mechanism
   \end{itemize}
   Contrary to the non-resonant case, the development of a secular model requires thus a two-step procedure. In Sec.~\ref{subsec:coo} and \ref{subsec:det}, we describe the new canonical coordinates used and the geometrical properties of the Hamiltonian function. Then, Section \ref{subsec:semisec} shows the transformation to an intermediary set of coordinates, referred here as ``semi-secular", in which the Hamiltonian is left with two degrees of freedom. The second change of coordinates (equivalent to a second averaging step) is described in Sec.~\ref{subsec:secres}: we finally obtain a one-degree-of-freedom secular system very similar to the non-resonant one. As previously, the phase portraits are preferentially drawn in some kind of secular elliptical elements (defined in Sec.~\ref{subsec:refcoo}), which are more directly interpretable than their canonical counterparts.
   
   \subsection{Coordinate change}\label{subsec:coo}
   In order to study the dynamics inside and around the $k_p\!:\!k$ resonance, we must at first isolate the resonant angle from the short periodic terms, as shown for instance in~\cite{MILANI-BACCILI_1998}. Basically, this consists in defining the angle $\sigma$ as a new canonical coordinate. From the Delaunay coordinates used so far (Eq.~\ref{eq:Delcoo}), this is done by a simple linear transformation applied to the angles:
   \begin{equation}\label{eq:rescoo}
        \begin{pmatrix}
           \sigma \\
           \gamma \\
           u \\
           v
        \end{pmatrix}
        =
        A
        \begin{pmatrix}
           l \\
           \lambda_p \\
           g \\
           h
        \end{pmatrix}
        =
        \begin{pmatrix}
           k & -k_p & k_p & k_p \\
           c & -c_p & c_p & c_p \\
           0 & 0 & 1 & 0 \\
           0 & 0 & 0 & 1
        \end{pmatrix}
        \begin{pmatrix}
           l \\
           \lambda_p \\
           g \\
           h
        \end{pmatrix}
   \end{equation}
   where $c$ and $c_p$ are integer coefficients, chosen such that:
   \begin{equation}
      \det A = c\,k_p-c_p\,k = 1
   \end{equation}
  This condition makes the transformation unimodular, so that any $2\pi$-periodic function with respect to the previous angles (as the Hamiltonian), is also $2\pi$-periodic with respect to the new ones. If we assume that $\sigma$ is a slow angle, it makes $\gamma$ the fastest circulating angle possible when $\lambda$ and $\lambda_p$ are related by~\eqref{eq:sigma}. In others words, $\gamma$ makes one revolution during a complete cycle of $\lambda$ and $\lambda_p$ ($k_p$ turns of $\lambda$ and $k$ turns of $\lambda_p$). Finally, note that we kept $\omega=g=u$ and $\Omega=h=v$ as independent coordinates, as we are interested in their secular evolutions. The transformation is then made canonical by applying $(A^T)^{-1}$ on the conjugated momenta:
   \begin{equation}
      \begin{pmatrix}
         \Sigma \\
         \Gamma \\
         U \\
         V
      \end{pmatrix}
      =
      \begin{pmatrix}
           -c_p & -c & 0 & 0 \\
            k_p &  k & 0 & 0 \\
           0 & 1 & 1 & 0 \\
           0 & 1 & 0 & 1
      \end{pmatrix}
      \begin{pmatrix}
         L \\
         \Lambda_p \\
         G \\
         H
      \end{pmatrix}
   \end{equation}
   and the coordinates $\{\lambda_{i\neq p}\}$ and $\{\Lambda_{i\neq p}\}$ remain unchanged. In these new variables, the Hamiltonian function $\mathcal{H}$ (Eq.~\ref{eq:Hgen}) rewrites:
   \begin{equation}
      \begin{aligned}
         \mathcal{H}\Big(\{\Lambda_{i\neq p}\},&\Sigma,\Gamma,U,V,\{\lambda_{i\neq p}\},\sigma,\gamma,u,v\Big) \\
         &= \mathcal{H}_0\Big(\{\Lambda_{i\neq p}\},\Sigma,\Gamma\Big) + \varepsilon\,\mathcal{H}_1\Big(\Sigma,\Gamma,U,V,\{\lambda_{i\neq p}\},\sigma,\gamma,u,v\Big) 
      \end{aligned}
   \end{equation}
   where the unperturbed part is:
   \begin{equation}
      \mathcal{H}_0 = -\frac{\mu^2}{2\,(k\,\Sigma+c\,\Gamma)^2} - n_p\,(k_p\,\Sigma+c_p\,\Gamma)+ \sum_{\substack{i=1\\i\neq p}}^{N}n_i\,\Lambda_i
   \end{equation}
   and the perturbation writes formally as in~\eqref{eq:Hfirst}:
   \begin{equation}
      \varepsilon\,\mathcal{H}_1 = - \sum_{i=1}^{N}\mu_i\left(\frac{1}{|\mathbf{r}-\mathbf{r}_i|} - \mathbf{r}\cdot\frac{\mathbf{r}_i}{|\mathbf{r}_i|^3}\right)
   \end{equation}
   However, the resonant part now behaves differently, because $r_p\equiv r_p(\sigma,\gamma,u,v)$ whereas for $i\neq p$ we have simply $r_i\equiv r_i(\lambda_i)$. In these coordinates, $\gamma$ is a fast angle, and $\sigma$ evolves with an intermediate (or ``semi-secular") time-scale.
   
   \subsection{Analytical development: details about the Hamiltonian function}\label{subsec:det}
   Before thinking of any new close-to-identity transformation, some general information can be grabbed about the resonant part of $\varepsilon\,\mathcal{H}_1$. Indeed, if we write the inverse of the mutual distances in terms of Legendre polynomials (Eq.~\ref{eq:devPL}), the angles $u=\omega$ and $v=\Omega$ appear in the perturbations only via the scalar product $\mathbf{r}\cdot\mathbf{r}_i$. With the planets on circular and coplanar orbits, it comes then:
   \begin{equation}
      \frac{\mathbf{r}\cdot\mathbf{r}_i}{r\,r_i} = \cos(\omega+\nu)\cos(\lambda_i-\Omega) + \sin(\omega+\nu)\sin(\lambda_i-\Omega)\cos I
   \end{equation}
   For the resonant planet $p$, that quantity writes in the new coordinates:
   \begin{equation}
      \frac{\mathbf{r}\cdot\mathbf{r}_p}{r\,r_p} = \cos(u+\nu)\cos(k\gamma-c\,\sigma+u)\ +\  \sin(u+\nu)\sin(k\gamma-c\,\sigma+u)\cos I
   \end{equation}
   where $\cos I$ should be replaced by:
   \begin{equation}
      \cos I = \frac{k_p\Sigma+c_p\Gamma+V}{k_p\Sigma+c_p\Gamma+U}
   \end{equation}
   and where the true anomaly $\nu$ is only function of $e$ and $M$, which write:
   \begin{equation}\label{eq:nueM}
      e = \sqrt{1-\left(\frac{k_p\Sigma+c_p\Gamma+U}{k\Sigma}\right)^2}
      \hspace{0.5cm}\text{and}\hspace{0.5cm}
      M = k_p\,\gamma - c_p\,\sigma
   \end{equation}
   The important point is that in the new coordinates, the resonant part of $\varepsilon\,\mathcal{H}_1$ is independent of the angle $v=\Omega$. Once again, this comes from our simple planetary model: in that case, the system ``particle + planet $p$" is invariant by rotation around the vertical axis.
   
   We can go further with some trigonometric identities:
   \begin{equation}
      \left\{
      \begin{aligned}
         2\cos(u+\nu)\cos(k\gamma-c\,\sigma+u) &= \cos(\nu+c\sigma-k\gamma)+\cos(\nu-c\sigma+k\gamma+2u) \\
         2\sin(u+\nu)\sin(k\gamma-c\,\sigma+u) &= \cos(\nu+c\sigma-k\gamma)-\cos(\nu-c\sigma+k\gamma+2u)
      \end{aligned}
      \right.
   \end{equation}
   which show that the resonant part of $\varepsilon\,\mathcal{H}_1$ is also $\pi$-periodic and symmetric with respect to $\pi/2$ in $u=\omega$.
   
   \subsection{Semi-secular Hamiltonian}\label{subsec:semisec}
   Thanks to our new definition of the angles~\eqref{eq:rescoo}, we can now safely switch to the ``semi-secular coordinates", for which the Hamiltonian is independent of the fast angles. The is done by the same close-to-identity transformation as we used in the non-resonant case. Thus, the semi-secular Hamiltonian writes:
   \begin{equation}
      \mathcal{K} = \mathcal{K}_0 + \varepsilon\,\mathcal{K}_1 + \mathcal{O}(\varepsilon^2)
   \end{equation}
   where $\mathcal{K}_0$ is functionally equal to $\mathcal{H}_0$ and $\varepsilon\,\mathcal{K}_1$ is functionally equal to the average of $\varepsilon\,\mathcal{H}_1$ with respect to the independent angles $\gamma$ and $\{\lambda_{i\neq p}\}$. At this point, it is interesting to note that, by mixing the old and new coordinates we have:
   \begin{equation}
      \gamma = \frac{1}{k_p}\lambda +\frac{1}{k_p}(c_p\,\sigma-u-v) = \frac{1}{k}\lambda_p +\frac{1}{k}(c\,\sigma-u-v)
   \end{equation}
   Hence, the average with respect to $\gamma$ is equivalent to an integral over $k_p$ turns of $\lambda$ (resp. $k$ turns of $\lambda_p$), expressing $\lambda_p$ (resp. $\lambda$) via the resonant angle~\eqref{eq:sigma}. Actually, this is the integral usually given for that kind of resonant problems \citep[see for instance][]{GALLARDO_2006b}, in which the coordinate change is just made implicit. Whatever the notation used, the semi-secular Hamiltonian (at first order of the planetary masses) writes formally:
   \begin{equation}
      \mathcal{K}\Big(\{\Lambda_{i\neq p}\},\Sigma,\Gamma,U,V,\sigma,u\Big) = \mathcal{K}_0\Big(\{\Lambda_{i\neq p}\},\Sigma,\Gamma\Big) + \varepsilon\,\mathcal{K}_1\Big(\Sigma,\Gamma,U,V,\sigma,u\Big)
   \end{equation}
   This time, we will not even try to obtain an analytical expression of $\mathcal{K}$, but the indications obtained from Sec.~\ref{subsec:det} are useful to understand its general form. In particular, the angle $v=\Omega$ has disappeared: indeed, the $i\neq p$ parts of $\varepsilon\,\mathcal{H}_1$ behave as in the non-resonant case (see Sec.~\ref{sec:nres}) and the $i=p$ part was \emph{already} independent of $v$. For the same reasons, $\mathcal{K}$ is also $\pi$-periodic and symmetric with respect to $\pi/2$ in $u=\omega$.
   
   The semi-secular constants of motion are then $V$, $\Gamma$ and the various $\{\Lambda_{i\neq p}\}$, and these lasts will now be omitted since they appear only as a constant term in $\mathcal{K}$. Concerning the $\Gamma$ momentum, one can notice that:
   \begin{equation}\label{eq:SUV}
      \left\{
      \begin{aligned}
         \Sigma &= \frac{1}{k}\sqrt{\mu a} - \frac{c}{k}\Gamma \\
         U &= \sqrt{\mu a}\left(\sqrt{1-e^2}-\frac{k_p}{k}\right) + \frac{1}{k}\Gamma \\
         V &= \sqrt{\mu a}\left(\sqrt{1-e^2}\cos I-\frac{k_p}{k}\right) + \frac{1}{k}\Gamma
      \end{aligned}
      \right.
   \end{equation}
   Considering that $\Gamma$ is now a constant, it can by seen as a free parameter of the transformation~\eqref{eq:SUV} from the semi-secular $(a,e,I)$ elements to the semi-secular $(\Sigma,U,V)$ momenta. The choice of $\Gamma$ being now only a matter of definition\footnote{We recall that the $\{\Lambda_i\}$ momenta were added artificially to the Hamiltonian to absorb its temporal dependence. Given that $\Gamma=k_pL+k\Lambda_p$, it is not surprising to get an entire liberty concerning its value.}, we will conveniently choose it equal to $0$. Finally, the semi-secular Hamiltonian function used in the following writes:
   \begin{equation}\label{eq:Kgen}
      \mathcal{K}\big(\Sigma,U,V,\sigma,u\big) = \mathcal{K}_0\big(\Sigma\big) + \varepsilon\,\mathcal{K}_1\big(\Sigma,U,V,\sigma,u\big)
   \end{equation}
   where:
   \begin{equation}
      \mathcal{K}_0\big(\Sigma\big) = -\frac{\mu^2}{2\,(k\Sigma)^2} - n_pk_p\Sigma
   \end{equation}
   and where $\varepsilon\,\mathcal{K}_1$ is obtained by computing numerically the required integrals, just as we did in Sec.~\ref{subsec:sanmod}. We are left with a two-degree-of-freedom system (the two angles being $\sigma$ and $u=\omega$), and several strategies can now be used to study its dynamics. The more general method is of course to compute Poincaré maps of the complete semi-secular system, but we did not find any example of it in the literature for transneptunian objects (although it would allow to detect a potential chaotic interaction between the two degrees of freedom). In our particular case, we will see that the intrinsic properties of the system allow to construct a more direct, secular representation.
   
   \subsection{Secular Hamiltonian}\label{subsec:secres}
   The method usually used in the literature for resonant secular models beyond Neptune is based on the crude model of~\cite{KOZAI_1985}. Indeed, in order to get immediate estimates of the resonant secular dynamics, Kozai chose to get rid of the extra degree of freedom by simply fixing $\Sigma$ and $\sigma$ at a supposed libration centre. Some authors, for better estimates, opted later for an assumed sinusoidal evolution of $\sigma$ with constant centre, frequency and amplitude \citep[see for instance][]{GOMES-etal_2005,GALLARDO-etal_2012}. Unfortunately, that kind of models is not adapted for the two following reasons: on the one hand, the choice of parameters (centre, frequency, amplitude) is problematic because we need an a priori knowledge of the dynamics. In particular we cannot choose an arbitrary libration centre: it must be an equilibrium point of the semi-secular Hamiltonian, otherwise the model is simply wrong... Since it is essential, then, to use a previous numerical integration, the secular model looses its utility as a tool to explore the variety of possible motions. On the other hand, these models just \emph{cannot} be considered as secular at all, because the oscillation parameters of $\sigma$ can actually vary a lot during the secular evolution. Therefore, the level curves obtained with such constant parameters are a very poor representation of the real trajectories, since they are valid only in a restricted neighbourhood of each point. That problem was recently mentioned by~\cite{BRASIL-etal_2014}: they picked up the oscillation parameters of $\sigma$ at different times from a numerical integration and plotted a collection of secular level curves, each graph being valid only at a time $t$ and in the very neighbourhood of the considered point. This is quite misleading, because different classes of dynamics seem to appear (as their so-called ``hibernating mode"), whereas they are actually just snapshots of an single global secular motion.
   
   Fortunately, we can also take advantage of the wide separation between the two time-scales associated with the two degrees of freedom in order to reduce the system to an integrable approximation. This technique is often called the ``adiabatic invariant approximation". Indeed, the experience shows that the oscillation period of $\sigma$ in that region ranges from a few tens of thousands to some million years (semi-secular time-scale), whereas the Lidov-Kozai cycles of $\omega$, as seen in the non-resonant case, are usually completed in more than a billion years (secular time-scale)\footnote{That separation prevents probably any occurrence of secondary resonance in our model.}. The method itself is not new: it was traditionally used to compute analytical proper elements for resonant or inclined asteroids, as in \cite{MORBIDELLI_1993}, \cite{LEMAITRE-MORBIDELLI_1994} or \cite{BEAUGE-ROIG_2001}. We can find it also in a series a paper devoted to the dynamics of asteroids in mean-motion resonance with Jupiter: see for instance \cite{WISDOM_1985}, \cite{MOONS-MORBIDELLI_1995} and \cite{MOONS-etal_1998}. On the following, the procedure is recalled and applied to our semi-secular Hamiltonian. Notice that we will not assume any particular evolution for $\sigma$ but accurately follow its variations.
   
   Our technique is based on two reference works: \cite{HENRARD_1993} which is a detailed course about the adiabatic invariant theory, and \cite{HENRARD_1990} where the useful transformation to action-angle coordinates is further detailed. For now, let us suppose that the dynamical system described by the semi-secular Hamiltonian~\eqref{eq:Kgen} is integrable. Let us also forget that it has two degrees of freedom but consider it as two \emph{independent} integrable systems, one for each pair of conjugated coordinates $(\Sigma,\sigma)$ and $(U,u)$. We will call $\nu_\sigma$ and $\nu_u$ the proper frequencies associated and assume that the resulting evolution of $u$ runs on a time-scale much larger than the one of $\sigma$, that is:
   \begin{equation}
      \xi = \frac{\nu_u}{\nu_\sigma}\ \ll\ 1
   \end{equation}
   If that relation holds, the action-angle coordinates $(J,\theta)$ related to the evolution of $(\Sigma,\sigma)$ for a fixed value of $(U,u)$ are a good approximation of the related ones in the complete two-degree-of-freedom system. More precisely, $J$ and $\theta$ are obtained up to order $\xi$. In particular, the momentum $J$ is not exactly conserved, but for a sufficiently small value of $\xi$ we can neglect its variations: in that case we say that $J$ is an ``adiabatic invariant" of the system. In the new coordinates, that we call secular, the Hamiltonian rewrites:
   \begin{equation}
      \mathcal{F}\big(J,U,V,\theta,u\big) = \mathcal{F}_0(J,U,V,u) + \mathcal{O}(\xi)
   \end{equation}
   where the new splitting is implicit and has nothing to do with the previous one (Eq.~\ref{eq:Kgen}). Following \cite{WISDOM_1985}, we will call $\mathcal{F}_0$ a ``quasi-integral" of motion. Neglecting the $\mathcal{O}(\xi)$ term, the dynamics can be described by the level curves of $\mathcal{F}$ in the $(U,u)$ plane: each point defines a one-degree-of-freedom subsystem with Hamiltonian $\mathcal{K}$ for $(U,u)$ fixed, and $J$ is the conserved action from the action-angle coordinates of that subsystem. In other words, the constant $J$ is related to a specific level curve of $\mathcal{K}$ in the $(\Sigma,\sigma)$ plane for $(U,u)$ fixed, called the ``guiding trajectory" by \cite{HENRARD_1993}. If we note $(\Sigma_0,\sigma_0)$ an arbitrary point of that level curve, the secular Hamiltonian neglecting the $\mathcal{O}(\xi)$ term is simply defined by:
   \begin{equation}\label{eq:Fsig}
      \mathcal{F}\big(J,U,V,u\big) = \mathcal{K}(\Sigma_0,U,V,\sigma_0,u)
   \end{equation}
   Note that no further averaging is required since the value of $\mathcal{K}$ is by definition the same all along the cycle. \cite{WISDOM_1985} used a similar representation to study the resonance $3\!:\!1$ with Jupiter in the planar problem\footnote{In \cite{WISDOM_1985}, take care that contrary to \cite{HENRARD_1993} or \cite{MILANI-BACCILI_1998}, the ``guiding trajectories" refers to the \emph{secular} time-scale, that is the level curves of $\mathcal{F}$ with respect to $(U,u)$.}. In addition, the method of ``fixing the slow variables by steps" was employed by~\cite{MILANI-BACCILI_1998} to describe the dynamics of Toro-type asteroids, but they did not used it to construct a secular model.
   
   Once the adiabatic invariance is postulated, the tricky part is to determine the action-angle coordinates of the one-degree-of-freedom subsystem. This can be done with the famous semi-analytical method of~\cite{HENRARD_1990}, as applied in the following \citep[see also][for an introduction]{LEMAITRE_2010}. Except from separatrices or equilibrium points, we can show that all the trajectories $(\Sigma\text{\small{(}}t\text{\small{)}},\sigma\text{\small{(}}t\text{\small{)}})$ for $(U,u)$ fixed are periodic, with a period $T_\sigma$ related to the level curve considered. Consequently, $2\pi/T_\sigma$ is the obvious proper frequency of the system, hence the choice of the new angle:
   \begin{equation}\label{eq:thdef}
      \theta = \nu_\sigma\,t + \theta_0 \text{\ \ \ \ with\ \ \ \ } \nu_\sigma = \frac{2\pi}{T_\sigma}
   \end{equation}
   Now, let us search for a complete canonical transformation of the form:
   \begin{equation}\label{eq:chgecoo}
      \begin{pmatrix}
         \Sigma  \\
         V \\
         \sigma \\
         v
      \end{pmatrix}
      =
      \begin{pmatrix}
         F(J,V',\theta) \\
         V' \\
         f(J,V',\theta) \\
         v' + \rho(J,V',\theta)
      \end{pmatrix}
   \end{equation}
   where $F$, $f$ and $\rho$ are $2\pi$-periodic functions of $\theta$. Note that we do not make any change to $U$ and $u$ because they are considered here as \emph{parameters}. In order to make~\eqref{eq:chgecoo} a canonical change of coordinates, three equations have now to be verified by the unknown functions $F$, $f$ and $\rho$. The first one writes:
   \begin{equation}
      1 = \frac{\partial f}{\partial \theta}\frac{\partial F}{\partial J} - \frac{\partial f}{\partial J}\frac{\partial F}{\partial \theta}
   \end{equation}
   and by two successive integrations by parts and applying the definition~\eqref{eq:thdef} of $\theta$, we get (apart from an arbitrary constant):
   \begin{equation}
      4\pi J = \int_0^{2\pi}\left(\frac{\partial f}{\partial \theta}F-\frac{\partial F}{\partial \theta}f\right)\mathrm{d}\theta = \int_0^{T_\sigma}\left(\dot{\sigma}\Sigma-\dot{\Sigma}\sigma\right)\mathrm{d}t
   \end{equation}
   or equivalently:
   \begin{equation}
      2\pi J = \frac{1}{2}\oint\left(\Sigma\,\mathrm{d}\sigma-\sigma\,\mathrm{d}\Sigma\right) = \oint\Sigma\,\mathrm{d}\sigma = -\oint\sigma\,\mathrm{d}\Sigma
   \end{equation}
   Except for the $2\pi$ factor, the new action $J$ is thus equal to a signed area, positive or negative according to the direction of motion along the level curve. In the case of oscillations around a central equilibrium, $2\pi J$ is the surface enclosed by the trajectory. On the contrary, it represents the area under the curve if $\sigma$ circulates \citep[see][for a simple example]{LEMAITRE_2010}. The two next equations enable to define the function $\rho$:
   \begin{equation}
      \frac{\partial \rho}{\partial \theta} = \frac{\partial f}{\partial V'}\frac{\partial F}{\partial \theta} - \frac{\partial f}{\partial \theta}\frac{\partial F}{\partial V'}
      \hspace{0.5cm}\text{;}\hspace{0.5cm}
      \frac{\partial \rho}{\partial J} = \frac{\partial f}{\partial V'}\frac{\partial F}{\partial J} - \frac{\partial f}{\partial J}\frac{\partial F}{\partial V'}
   \end{equation}
   and by direct integration and a judicious choice of origin for $\theta$, we get simply:
   \begin{equation}
      \rho(J,V',\theta) = \int_0^\theta\left(\frac{\partial f}{\partial V'}\frac{\partial F}{\partial \theta} - \frac{\partial f}{\partial \theta}\frac{\partial F}{\partial V'}\right)\mathrm{d}\theta
   \end{equation}
   Concerning the constant frequency of $v'$, it is straightforward to get it from the change of coordinates~\eqref{eq:chgecoo}:
   \begin{equation}
      \nu_v = \frac{\mathrm{d}v'}{\mathrm{d}t} = \frac{\mathrm{d}v}{\mathrm{d}t} - \frac{\mathrm{d}\rho}{\mathrm{d}t}
   \end{equation}
   and by integration between $0$ and $T_\sigma$ we have simply:
   \begin{equation}
      \nu_v = \frac{v(T_\sigma) - v(0)}{T_\sigma}
   \end{equation}
   In practice, since the dynamics of $v=\Omega$ is well decoupled from $\sigma$ (just as for $u$), the function $\rho$ will be only a little correction, that is $v'\approx v$. Anyway, its calculation is required only if we are interested in the temporal evolution of $\Omega$ as a function of the new coordinates.
   
   Naturally, the coordinate change~\eqref{eq:chgecoo} is only implicit, since neither $F$ nor $f$ have an explicit definition. Nevertheless, the correspondence between $(\Sigma,V,\sigma,v)$ and $(J,V',\theta,v')$ can be realized numerically by integrating the equations of motion defined by the semi-secular Hamiltonian $\mathcal{K}$ for $(U,u)$ fixed. Indeed, once we know the period $T_\sigma$ and the functions $\Sigma(t)$, $\sigma(t)$ and $v(t)$ for a chosen value of $J$, the link toward $\theta(t)$ and $v'(t)$ is straightforward for all $t$: the coordinate change is simply defined by identification. In our case, since we are only interested in the value of the secular Hamiltonian $\mathcal{F}(J,U,V,u)$, the procedure is the following:
   \begin{enumerate}
      \item Choose a behaviour for $\sigma$: oscillation or circulation (because the definition of $J$ differs from a case to the other).
      \item Choose the parameters $J$ and $V$.
      \item For each point $(u,U)$ where we want the value of~$\mathcal{F}$, do:
      \begin{enumerate}
         \item On the $(\Sigma,\sigma)$ plane, look for the equilibrium point(s) of $\mathcal{K}$ with $(U,u)$ as constants. This is done numerically with minimization/maximization routines.
         \item Look also for the position of the separatrix, in order to define the boundaries of the search.
         \item In the domain of interest (inside or outside the separatrix, see point~1), search for the level curve corresponding to an area $A=2\pi J$. This is done by integrating numerically the semi-secular equations of motion for $(U,u)$ fixed, and applying a Newton algorithm with respect to the initial conditions. Indeed, the surface over time can be added among the dynamical equations:
         \begin{equation}
            \dot{A} = \frac{1}{2}\left(\dot{\sigma}\Sigma-\dot{\Sigma}\sigma\right)
         \end{equation}
         with another Newton algorithm or the method of~\cite{HENON_1982} to stop the integration exactly after a complete cycle.
         \item If there is no trajectory with the required surface in the domain (for instance if the separatrices are too narrow to contain it), stop with a warning: that combination of parameters is impossible. Conversely if a correct initial condition $(\Sigma_0,\sigma_0)$ has been found, pick up the period $T_\sigma$ associated to verify that it is well below the secular time-scale. Some additional output can also be printed (position of the equilibrium point(s), width of the separatrices...).
         \item The value of the secular Hamiltonian $\mathcal{F}(J,U,V,u)$ is finally given by~\eqref{eq:Fsig}.
      \end{enumerate}
   \end{enumerate}
   Practically, the computation of the semi-secular Hamiltonian $\mathcal{K}$ and its partial derivatives (for the iterative numerical integrations) is rather CPU-time consuming because it always implies the numerical averaging over the short periods (see Sec.~\ref{subsec:semisec}). Following the idea of~\cite{LEMAITRE-MORBIDELLI_1994}, we thus perform a 2D cubic splines interpolation of $\mathcal{K}$ in the $(\Sigma,\sigma)$ plane around the equilibrium point(s) (between steps~3b and~3c). The partial derivatives are then calculated by direct derivation of the splines and the numerical integration is performed with virtually no cost. Finally, the computation of a complete map is easily parallelized since each point is independent of the others. Naturally, we can also speed up the computation by reducing the resolution.
   
   \subsection{Reference coordinates}\label{subsec:refcoo}
   We are now able to draw the level curves of the secular Hamiltonian $\mathcal{F}$ in the $(U,u)$ plane with respect to the two fixed parameters $V$ and $J$. However, it would be convenient to express it with coordinates more directly meaningful, as we did in the non-resonant case. First of all, let us define a reference semi-major axis $a_0$ (it's choice, somehow arbitrary, is discussed later). Since the momentum $V$ is a secular constant of motion, we have:
   \begin{equation}
      V = \sqrt{\mu a} \left(\eta - k_p/k\right) = \text{const.}
   \end{equation}
   where we wrote $\eta = \sqrt{1-e^2}\cos I$. The constant $V$ can then be replaced by the parameter:
   \begin{equation}
      \eta_0 = \frac{V}{\sqrt{\mu a_0}} + \frac{k_p}{k}
   \end{equation}
   Also, the variable $U$ can be replaced by a reference perihelion $\tilde{q} = a_0(1-\tilde{e})$, where the reference eccentricity is defined by:
   \begin{equation}
      \tilde{e}\,^2 = 1-\left(\frac{U}{\sqrt{\mu a_0}}+\frac{k_p}{k}\right)^2
   \end{equation}
   At this point, one can remark that $a_0$ should be chosen big enough to allow a constant $\eta_0\in[-1,1]$ and a positive value for $\tilde{e}\,^2$. Finally, we can also define a reference inclination by setting:
   \begin{equation}
      \sqrt{1-\tilde{e}\,^2}\cos\tilde{I} = \eta_0
   \end{equation}
   The plane $(\tilde{q},\omega)$ is thus entirely equivalent to the plane $(U,u)$, and the parameter $\eta_0$ is entirely equivalent to the $V$ constant. The point is now to determine if these new quantities have a physical meaning, and to what extend they represent the real secular orbit of the particle. Actually, we can verify (see Sec.~\ref{sec:ex}) that the secular variations of the semi-major axis are always rather small, such that it is never far from a central approximate value. If such a value is chosen for $a_0$, the function $\eta(t)$ will always remain close to the constant $\eta_0$, and we will also have $\tilde{e}(t) \approx e(t)$ and $\tilde{q}(t) \approx q(t)$. The parameter $\eta_0$ acts then as the Kozai constant of the non-resonant case, linking the inclination and the eccentricity (even if this time, it is only in an approximative way). Consequently, in all what follows the level curves of the secular Hamiltonian $\mathcal{F}$ will be plotted in the $(\tilde{q},\omega)$ plane with $\eta_0$ as parameter. Naturally, the value of $a_0$ chosen will be given to let us recover the original canonical coordinates $U$ and $V$.
   
   Concerning the parameter $J$, its link with the elliptical elements is so abstract that we will not try to redefine it. Let us just keep in mind that its value is always negative if $\sigma$ librates (as in our case, the equilibriums are maxima), and that its magnitude is related to the enclosed area in the $(a,\sigma)$ plane, that is to the amplitude of oscillation.

\section{Application and examples}\label{sec:ex}
   This last section presents some examples of use of the resonant secular model. A variety of typical cases are provided to emphasis the main vantages and limitations of the method. As a quick check, the secular model will also be confronted to numerical integrations of the osculating and semi-secular systems\footnote{For integrating numerically the semi-secular system, the required partial derivatives of $\mathcal{K}$ are obtained by inverting the derivative and integration symbols in the expression of $\varepsilon\,\mathcal{K}_1$. Some nested derivatives can become a bit cumbersome: do not forget, for instance, that the true anomaly is function of $e$ and $M$, themselves functions of $\Sigma$, $U$, $\gamma$ and $\sigma$ via~\eqref{eq:nueM}.}. Section~\ref{subsec:sing} presents the ideal case, where the adiabatic invariant $J$ is well defined all over the surface $(\omega,\tilde{q})$ considered. In Sec.~\ref{subsec:sepcross}, we show that a secular description is still possible for higher values of $|2\pi J|$ even if $\sigma$ switches from oscillation to circulation. Finally, Section~\ref{subsec:double} illustrates the most complex case in that region, where the existence of two deforming resonance islands leads necessarily to a discontinuity in the secular phase portraits.

   \subsection{Single resonance island and small values of $J$}\label{subsec:sing}
\begin{figure}
      \centering
      \includegraphics[width=0.8\linewidth]{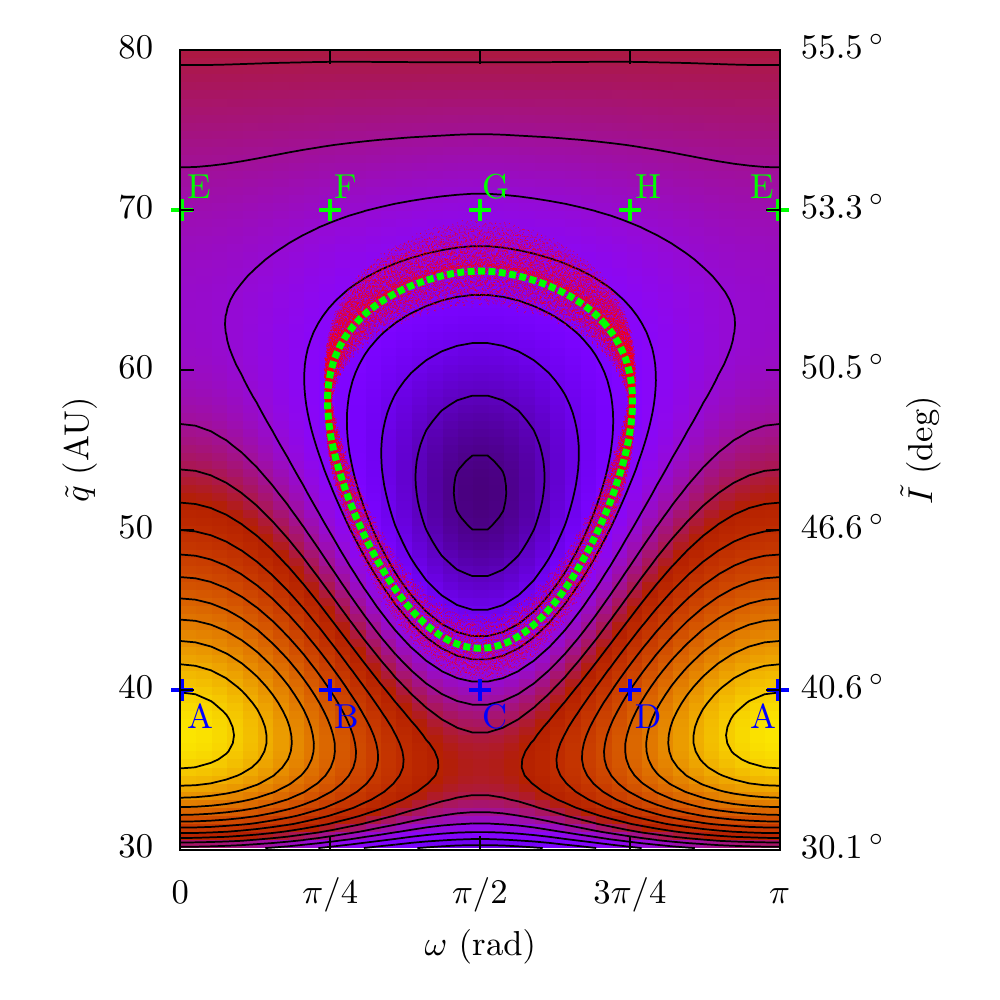}
      \caption{Level curves of the secular Hamiltonian $\mathcal{F}$ for the resonance $2\!:\!37$ with Neptune.  The parameters are $\eta_0 = 0.44$ and $2\pi J=-2.6\times 10^{-4}$ AU$^2$rad$^2/$yr. To define $\eta_0$ and construct the vertical axis, the reference semi-major axis chosen is $a_0=210.9944$ AU (see Fig.~\ref{fig:Fsec_gen_param} where that value is obvious). See text for the symbols.}
      \label{fig:Fsec_gen}
   \end{figure}

\begin{sidewaysfigure}
      \begin{minipage}[c][12cm][]{\textwidth}\end{minipage}
      \vfill
      \begin{minipage}[c]{\textwidth}
         \centering
         \includegraphics[width=\linewidth]{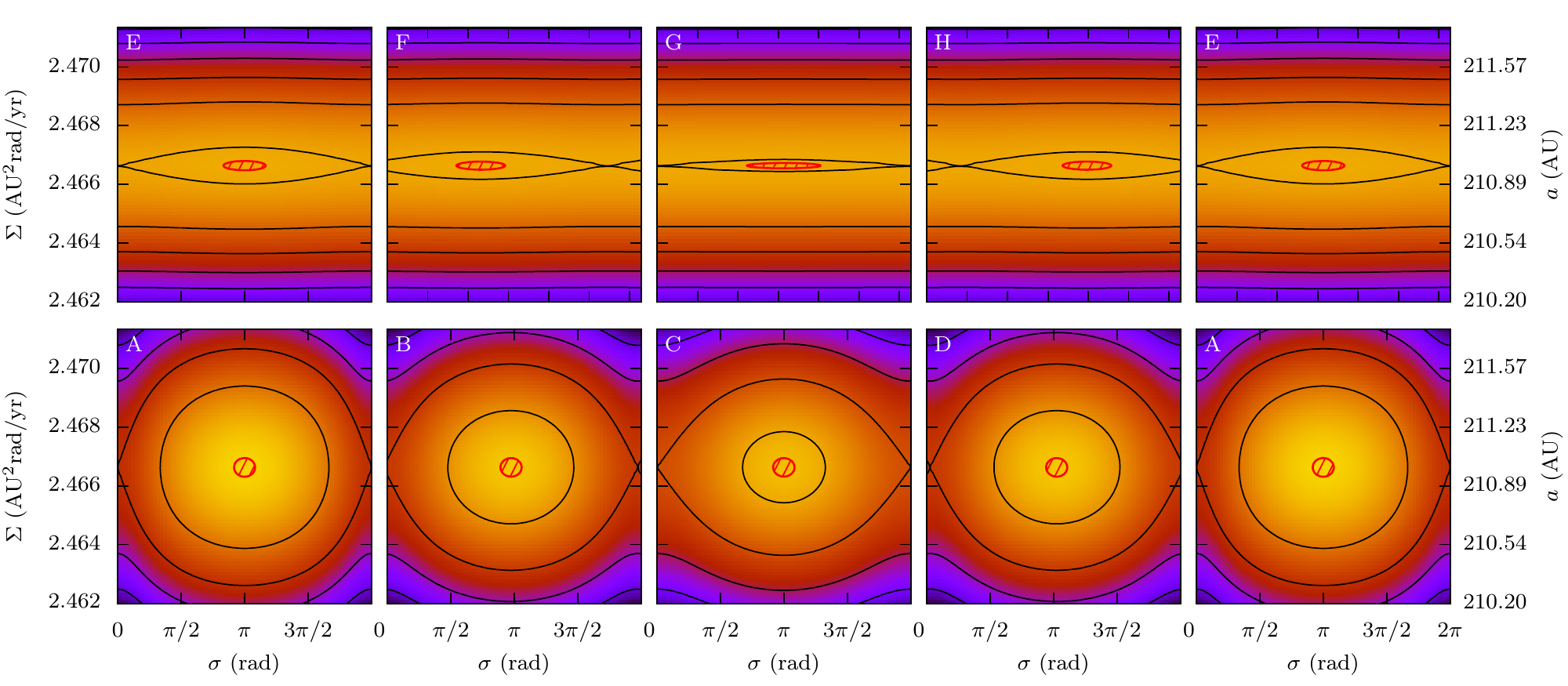}
         \captionof{figure}{Level curves of the semi-secular Hamiltonian $\mathcal{K}$, for $(U,u)$ fixed according to the points A-H of Fig.~\ref{fig:Fsec_gen}. The trajectory enclosing the surface $2\pi J$ is shown in red and the semi-major axis corresponding to $\Sigma$ is given on the right. The centre is perfectly at $\sigma=\pi$ for the points (A,C,E,G) but slightly shifted for (B,F) and (D,G) symmetrically on the left and on the right. For the points E-H, the $\Sigma$-width of the resonance island is very narrow because of the high perihelion (see Fig.~\ref{fig:Fsec_gen}), which makes the red surface to flatten.}
         \label{fig:ssgen4070}
      \end{minipage}
   \end{sidewaysfigure}

   Let us begin with the simplest case, that is when the semi-secular plane $(\Sigma,\sigma)$ contains a single island of resonance. Fortunately, this is \emph{almost}\footnote{For instance, we found that the resonance $2\!:\!11$ with Neptune has a double island if $\eta_0=-0.65$, with $\omega\sim\pi/2$ and $\tilde{q}\sim 34$ AU (where we chose $a_0=93.9872$ AU). However, the required range of parameters is very narrow.} always the case for exterior mean-motion resonances other than type $1\!:\!k$ \citep[see][for more details]{GALLARDO_2006b}. Of course, that single island will possibly deform and shift a lot during the secular evolution of $(U,u)$, but the secular dynamics is well defined as long as the surface enclosed by the separatrix remains greater than $2\pi J$. Figure~\ref{fig:Fsec_gen} shows an example of level curves obtained for such a case (black lines). As this is the first graph, some extra information is provided to recall the different time-scales and appreciate the efficiency of the method:
   \begin{enumerate}
      \item The little red dots come from a complete numerical integration (osculating elements): the equations are given by the initial Hamiltonian $\mathcal{H}$~\eqref{eq:Hgen} without any approximation. The fast angles make the plot somewhat messy, mainly because of the shift of the Solar System barycentre.
      \item The dashed green line is the result of a numerical integration of the semi-secular system: the equations are given by the two-degree-of-freedom semi-secular Hamiltonian $\mathcal{K}$~\eqref{eq:Kgen}, that is after removing the short periodic terms from $\mathcal{H}$. The curve follow very well the average pattern of the red dots and the oscillations due to the second degree of freedom are smaller than the curve width. See Fig.~\ref{fig:ssevol} for a detailed output of that numerical integration (in particular we can see that the cycle is completed in about $1.12$~Gyrs).
      \item Finally, the colour shades show the value of the one-degree-of-freedom secular Hamiltonian $\mathcal{F}$~\eqref{eq:Fsig}. Each point is obtained from the action-angle coordinates of $\mathcal{K}$ assuming the adiabatic invariance. The secular dynamics is then given by the level curves of $\mathcal{F}$ (black contours).
   \end{enumerate}
   In order to illustrate the passage from the semi-secular to the secular coordinates, Figure~\ref{fig:ssgen4070} shows the level curves of the semi-secular Hamiltonian~$\mathcal{K}$ corresponding to ten points of Fig.~\ref{fig:Fsec_gen} (letters). The level curve that encloses the required area defines the value of the secular Hamiltonian~$\mathcal{F}$. For that set of parameters, the surface $|2\pi J|$ is sufficiently small to fit easily inside the separatrix but its contours can be rather distorted. In particular, the narrowing of the $\Sigma$-width of the island, when the perihelion increases, forces $\sigma$ to oscillate with a larger amplitude. By the way, note that the general properties of $\mathcal{K}$ in $\omega$ are easily recognizable: $\pi$-periodicity and symmetry with respect to $\pi/2$ (see Sec.~\ref{subsec:semisec}).
      
\begin{figure}
      \includegraphics[width=\textwidth]{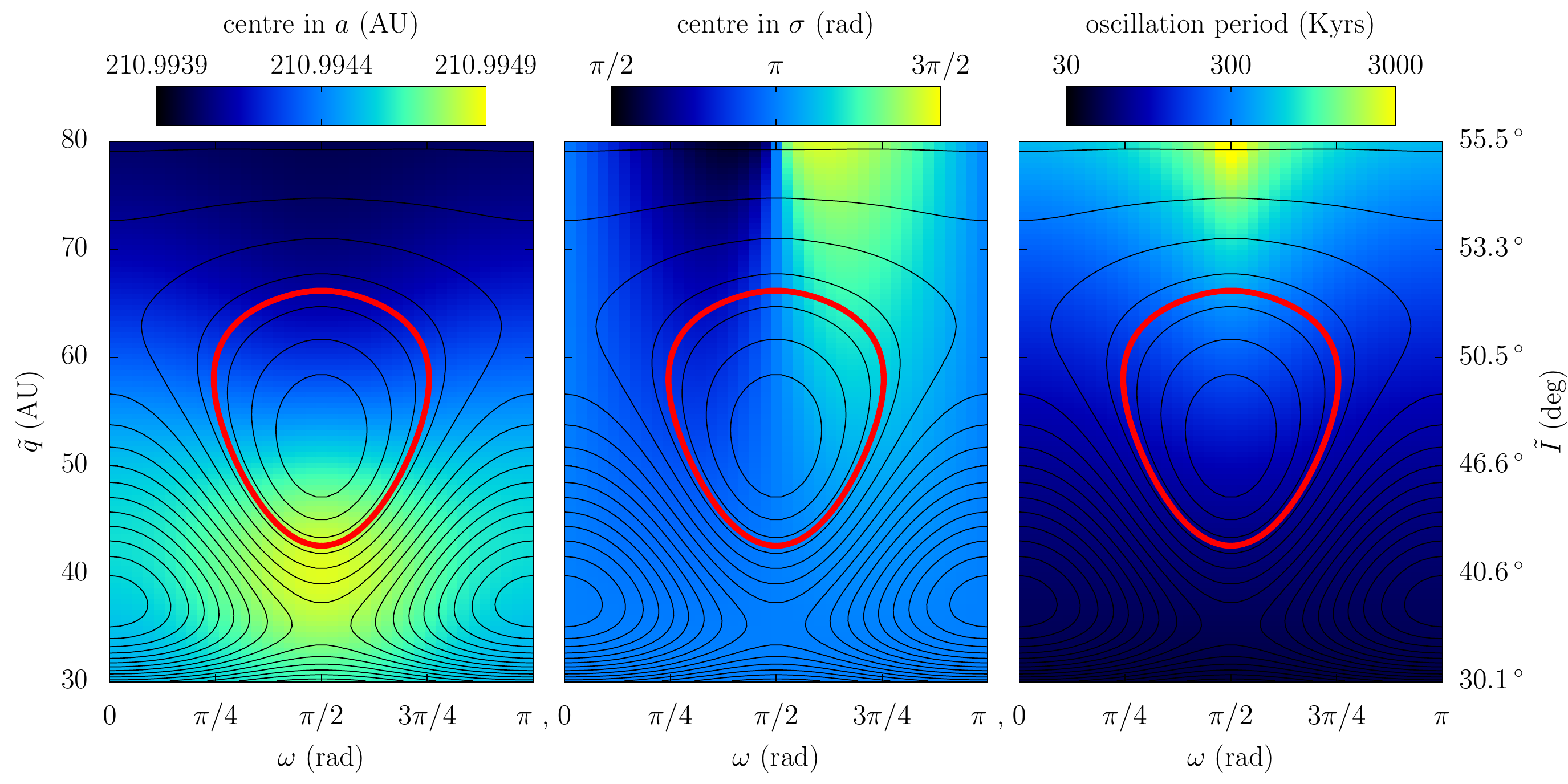}
      \caption{The level curves of Fig.~\ref{fig:Fsec_gen} are plotted in front of some characteristics of the resonance island in the plane $(\Sigma,\sigma)$ used to get the action-angle coordinates of~$\mathcal{K}$. On the left graphic, the semi-major axis is used instead of $\Sigma$ for a more direct interpretation. The middle plot shows that in that particular case, the oscillation centre of $\sigma$ oscillates itself around $\pi$. On the right graphic, the oscillation period refers to the trajectory enclosing the required area $2\pi J$: even if it varies a lot (note the log-scale), it remains much smaller than the Giga-year secular periods. The red line represents the result of a numerical integration of the semi-secular system (the same as the green dashed line of Fig.~\ref{fig:Fsec_gen}).}
      \label{fig:Fsec_gen_param}
   \end{figure}

\begin{figure}
      \includegraphics[width=\textwidth]{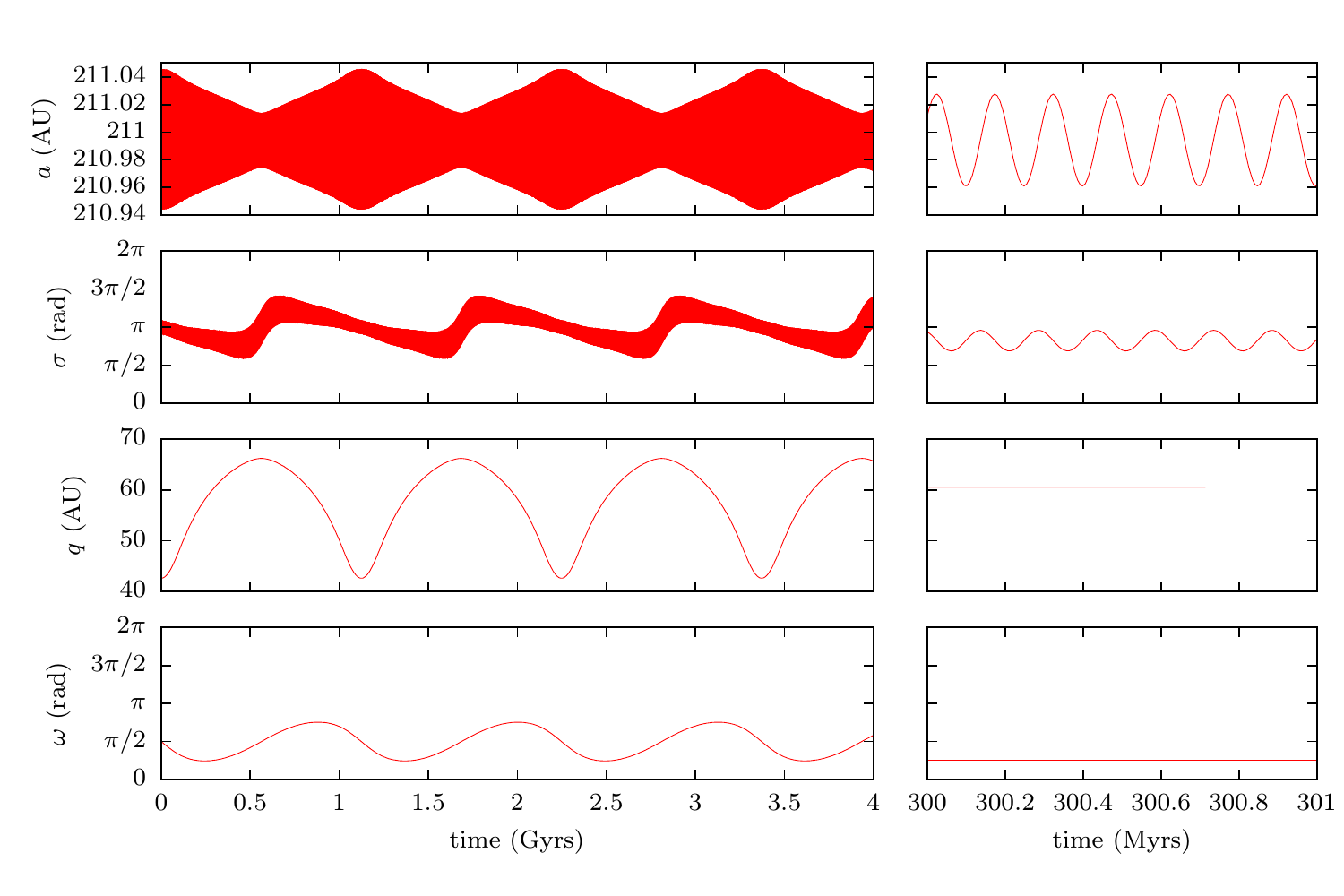}
      \caption{Numerical integration of the two-degree-of-freedom semi-secular system. That trajectory corresponds to the green dashed line of Fig.~\ref{fig:Fsec_gen} and the red line of Fig.~\ref{fig:Fsec_gen_param}. The semi-major axis is given instead of $\Sigma$ and the perihelion instead of $U$ (see Eq.~\ref{eq:SUV} for the correspondence). On the right, an enlargement underlines the two-time-scaled dynamics (the small oscillations of $q$ and $\omega$ are hidden in the curve width).}
      \label{fig:ssevol}
   \end{figure}

   Figure~\ref{fig:Fsec_gen_param} presents the same level curves as Fig.~\ref{fig:Fsec_gen}, but with the position of the centre of the resonance island on background shades, as well as the period of oscillation. The amplitudes are not shown here, but Figure~\ref{fig:ssgen4070} gives an idea of their variations. Following a particular level curve, we can see the important changes of the oscillation parameters underwent by the particle (the red line and Figure~\ref{fig:ssevol} give a specific example of it). This invalidates any secular model for which the resonance angle is supposed fixed or sinusoidal. Nevertheless, the central value of the semi-major axis is indeed rather stable: it is actually imposed by Neptune's semi-major axis. This justifies the use of ``reference coordinates" as a short-cut from the secular variable $U$ to the secular orbital elements $e$ and $I$ (see Sec.~\ref{subsec:refcoo}).
      
\begin{figure}
      \centering
      \includegraphics[width=0.6\textwidth]{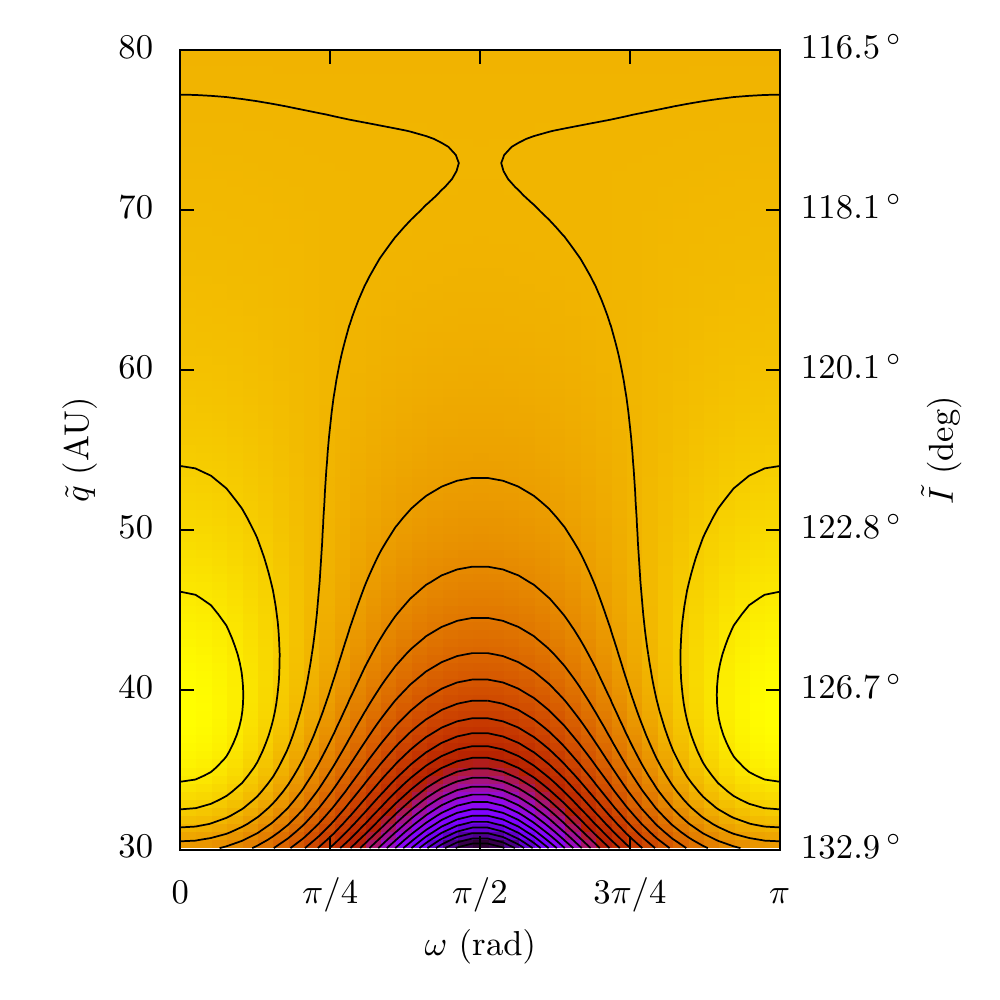}
      \caption{Level curves of the secular Hamiltonian $\mathcal{F}$ for the resonance $2\!:\!37$ with Neptune (reference semi-major axis chosen: $a_0=210.9944$ AU). The parameters are $\eta_0 = -0.35$ and $2\pi J=-1.7\times 10^{-5}$ AU$^2$rad$^2/$yr. Note that these orbits are retrograde.}
      \label{fig:Fsec_gen2}
   \end{figure}

   Finally, Figure~\ref{fig:Fsec_gen2} gives another example of secular dynamics with a small value of $|2\pi J|$. The resonance is the same as Fig.~\ref{fig:Fsec_gen} but another set of parameters is chosen: one can notice the extreme richness of possible behaviours, with many different ways to raise the perihelion distance. However, it is a general result that the $\Sigma$-width of the resonance island becomes much wider when the perihelion tends to the semi-major axis of Neptune. Since this is also the case for all the neighbouring resonances, we must keep in mind that for grazing orbits the overlapping of resonances can introduce some chaos and push the particle out of the resonance considered. This happens indeed for the largest trajectories of Fig.~\ref{fig:Fsec_gen2} but their major portions, though, are perfectly regular (as shown by various numerical integrations of the unaveraged system). To fix ideas, the biggest cycle represented is completed in about $40$ Gyrs, where more than $32$ Gyrs are spent with $\tilde{q}>70$ AU.
      
   \subsection{Separatrix crossings}\label{subsec:sepcross}
   For high values of the perihelion distance, we saw that the $\Sigma$-width of the resonance island becomes very small (see Fig.~\ref{fig:ssgen4070}). This has a stabilizing effect because the various resonances become very isolated from each other (no overlapping), but what if the island becomes so narrow that the area $|2\pi J|$ cannot fit inside any more? From a technical point of view, the values of the parameters are simply \emph{incompatible}, so what if a secular level curve \emph{leads} the particle to such a region? The resulting trajectory can be described as follows: the semi-secular separatrices in the $(\Sigma,\sigma)$ plane come closer and closer to the trajectory, making raise the amplitude of oscillation of $\sigma$, along with a drastic enlargement of its period. Then, the particle can spend some time near the unstable equilibrium point, breaking the adiabatic invariant hypothesis. Fortunately, this ``freeze" is usually quite short because $U$ and $u$ are still evolving. Hence, the particle is simply pushed outside of the resonance island and $\sigma$ begins to circulate. On can remember that the method applied in Sec.~\ref{subsec:secres} is also valid for circulation\footnote{The proximity of the resonance still invalidates a fully non-resonant secular model.}, but the geometrical definition of $J$ has to be changed. Consequently, the only way to handle the crossing in a secular way is to change model: the secular trajectory is then defined by parts, each of them being quasi-integrable. For a given trajectory, the problem is now to link the segments. There is actually no way to deduce the exact value of the new $J$ constant adopted by the system, because it depends of the precise position of the particle when the separatrix crossing occurs. On a secular time-scale, this can be seeing as a random transition \citep[see][and references therein]{HENRARD_1993}. In particular, since in our case the island is quasi symmetric on the $\Sigma$ axis, there is roughly $50\%$ of probability to begin circulate toward the left (above the island) or the right (under the island). However, if the new secular level curve is periodic the particle is bound to re-enter the resonance in a configuration similar to when it left it. After the new separatrix crossing, the value of $J$ will thus be approximatively restored (apart from some chaotic diffusion).
   
   That mechanism was described thoroughly by \cite{WISDOM_1985} in the case of the resonance $3\!:\!1$ with Jupiter and the associated Kirkwood gap. Near the discontinuities of the secular Hamiltonian (that is when the crossings occur in the semi-secular system), he named "zone of uncertainty" the region in which the adiabatic invariant hypothesis is invalidated. In his model, any passage through this zone produced a jump at possibly planet-crossing eccentricities. Moreover, even if the particle re-entered the resonance afterwards, the value of the adiabatic invariant was not recovered, which produced a large-scale chaotic behaviour. He pointed out that that kind of chaos is \emph{not} due to a mean-motion resonance overlap (that is a short time-scale effect), contrary to many chaotic orbits of asteroids observed in the Solar System. It could be explained, though, by an overlap of secondary resonances between $\sigma$ and $\omega$ which happen to have comparable frequencies of oscillation/circulation in these regions. Subsequently, \cite{NEISHTADT_1987} developed rather general methods to trace the evolution of the adiabatic invariant near and during such discontinuities. In particular, their application to the problem of \cite{WISDOM_1985} results in a probabilistic model governing the new value of the invariant when the particle re-enters the resonance.
   
   Fortunately, the orbits described in the present paper are much more regular and predictable because the separation between the two time-scales is much larger. This was quite visible on Fig.~\ref{fig:ssevol}, where it is impossible to resolve the two time-scales with a single time unit. This implies that the "zone of uncertainty" is extremely narrow in our problem: on a secular time-scale, it is crossed quasi-instantaneously. Hence, since there is almost never any interaction between the two degrees of freedom, the new value of $J$ is very predictable for each possible transition.
   
\begin{figure}
      \centering
      \includegraphics[width=\textwidth]{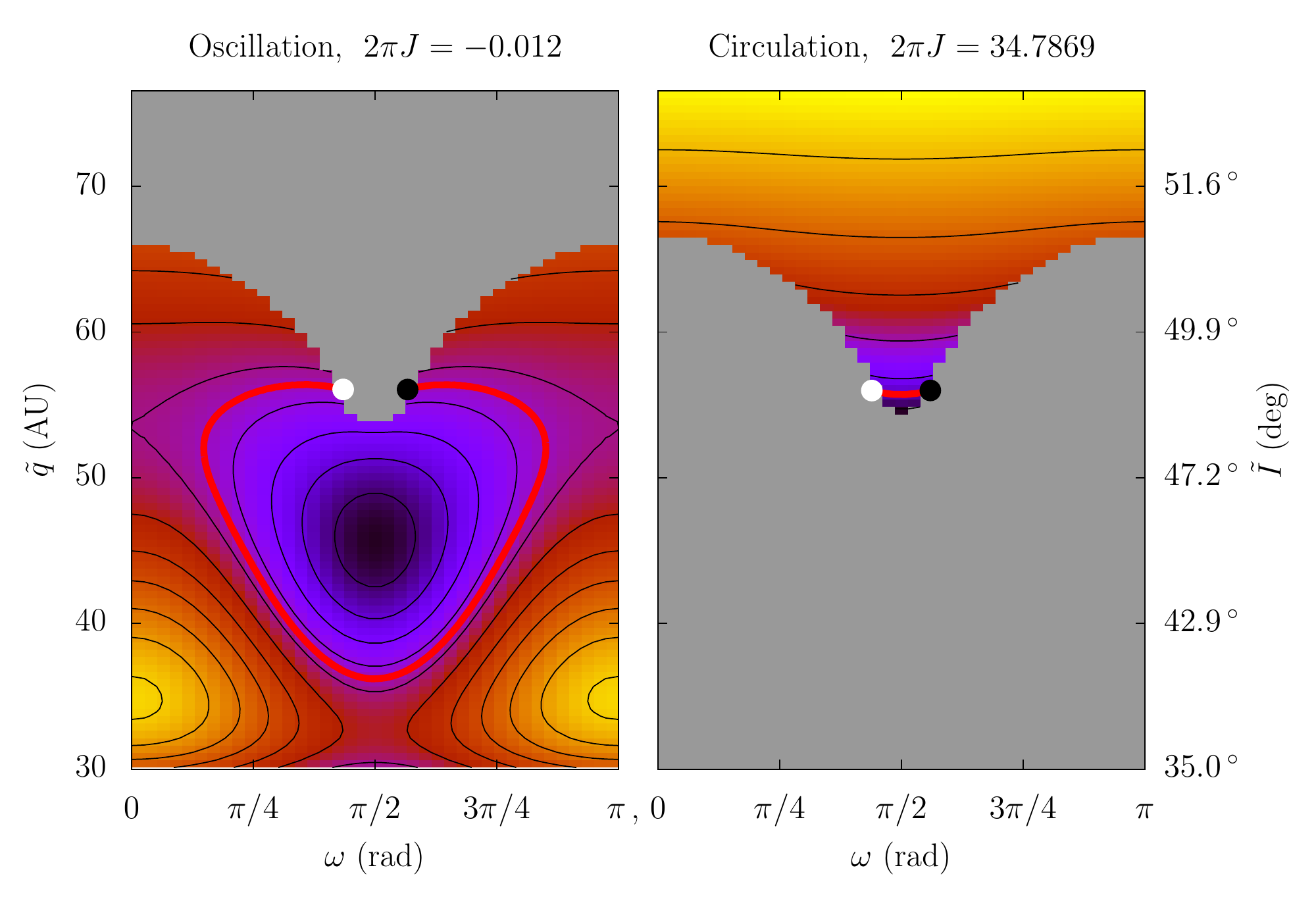}
      \caption{Level curves of the secular Hamiltonian $\mathcal{F}$ for the resonance $2\!:\!11$ with Neptune (reference semi-major axis chosen: $a_0=93.9872$ AU). The common parameter is $\eta_0 = 0.6$ and $J$ is given above the graphs in AU$^2$rad$^2/$yr. The red trajectory passes from a secular model to the other according to the colour spots (white-white, then black-black).  The saw teeth of the background colour are due to the resolution.}
      \label{fig:ssCross1}
   \end{figure}

\begin{figure}
      \centering
      \includegraphics[width=\textwidth]{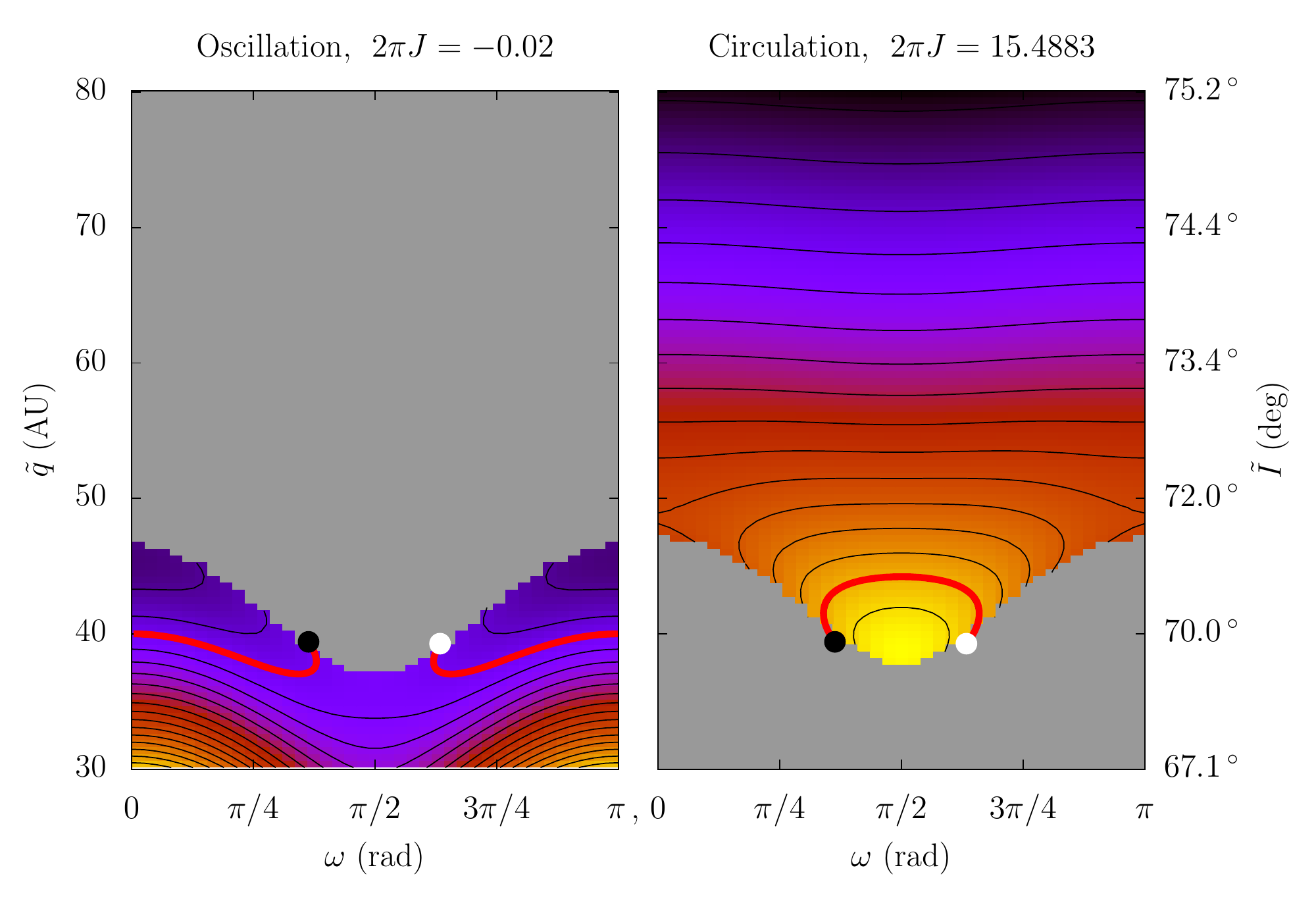}
      \caption{Level curves of the secular Hamiltonian $\mathcal{F}$ for the resonance $2\!:\!37$ with Neptune (reference semi-major axis chosen: $a_0=210.9944$ AU). The common parameter is $\eta_0 = 0.2$ and $J$ is given above the graphs in AU$^2$rad$^2/$yr. The red trajectory passes from a secular model to the other according to the colour spots (white-white, then black-black). The saw teeth of the background colour are due to the resolution.}
      \label{fig:ssCross2}
   \end{figure}

\begin{figure}
      \centering
      \includegraphics[width=\textwidth]{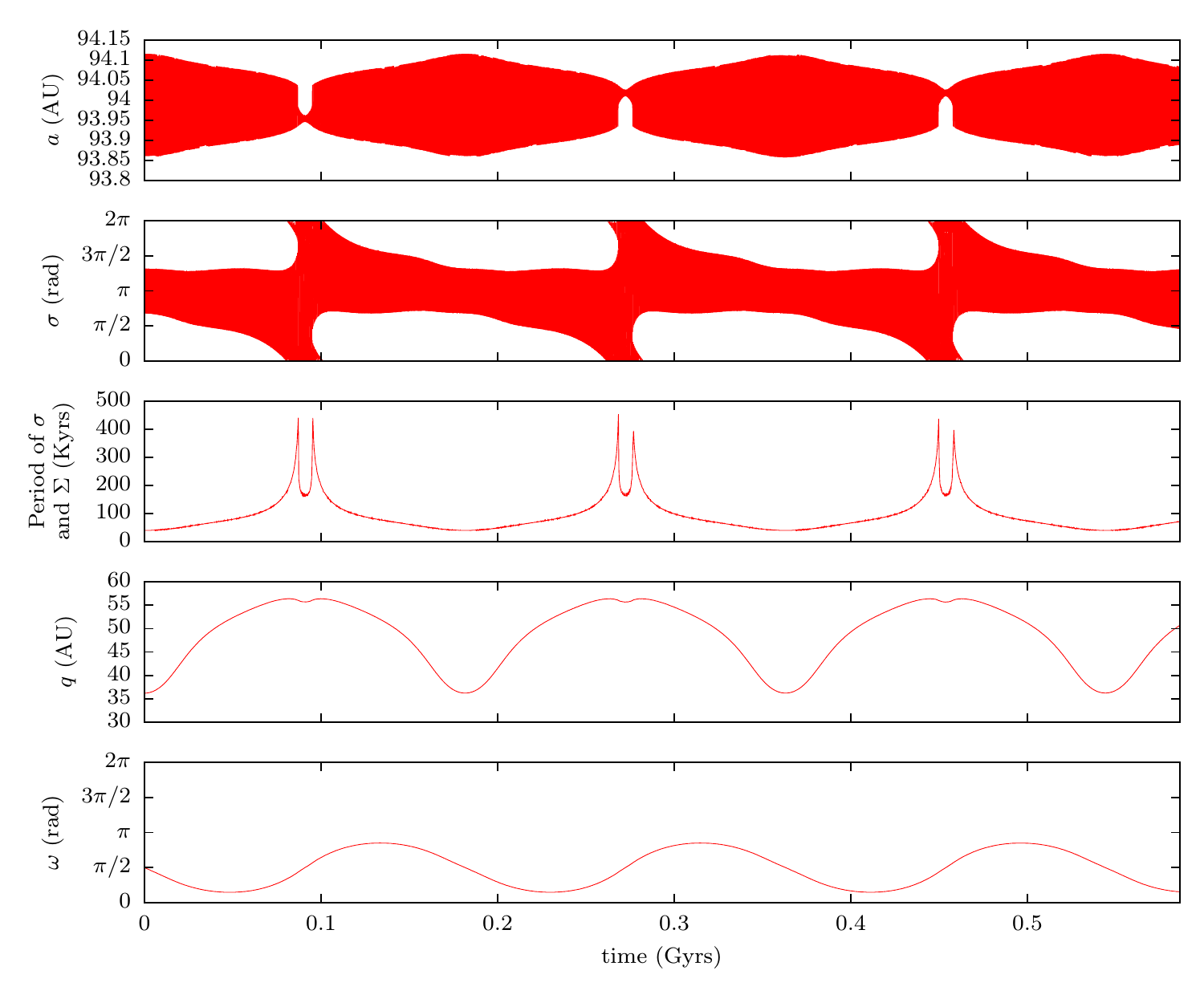}
      \caption{Numerical integration of the two-degree-of-freedom semi-secular system. That trajectory corresponds to the red line of Fig.~\ref{fig:ssCross1}. The semi-major axis is given instead of $\Sigma$ and the perihelion instead of $U$ (see Eq.~\ref{eq:SUV} for the correspondence). The period of $\sigma$ (oscillation/circulation) is given on the middle graph, where the separatrix crossings are obvious (the period tends to infinity). The first circulation phase is toward the right (see $a$ and $\sigma$), whereas the second and the third are toward the left.}
      \label{fig:ssCross1-details}
   \end{figure}

\begin{figure}
      \centering
      \includegraphics[width=\textwidth]{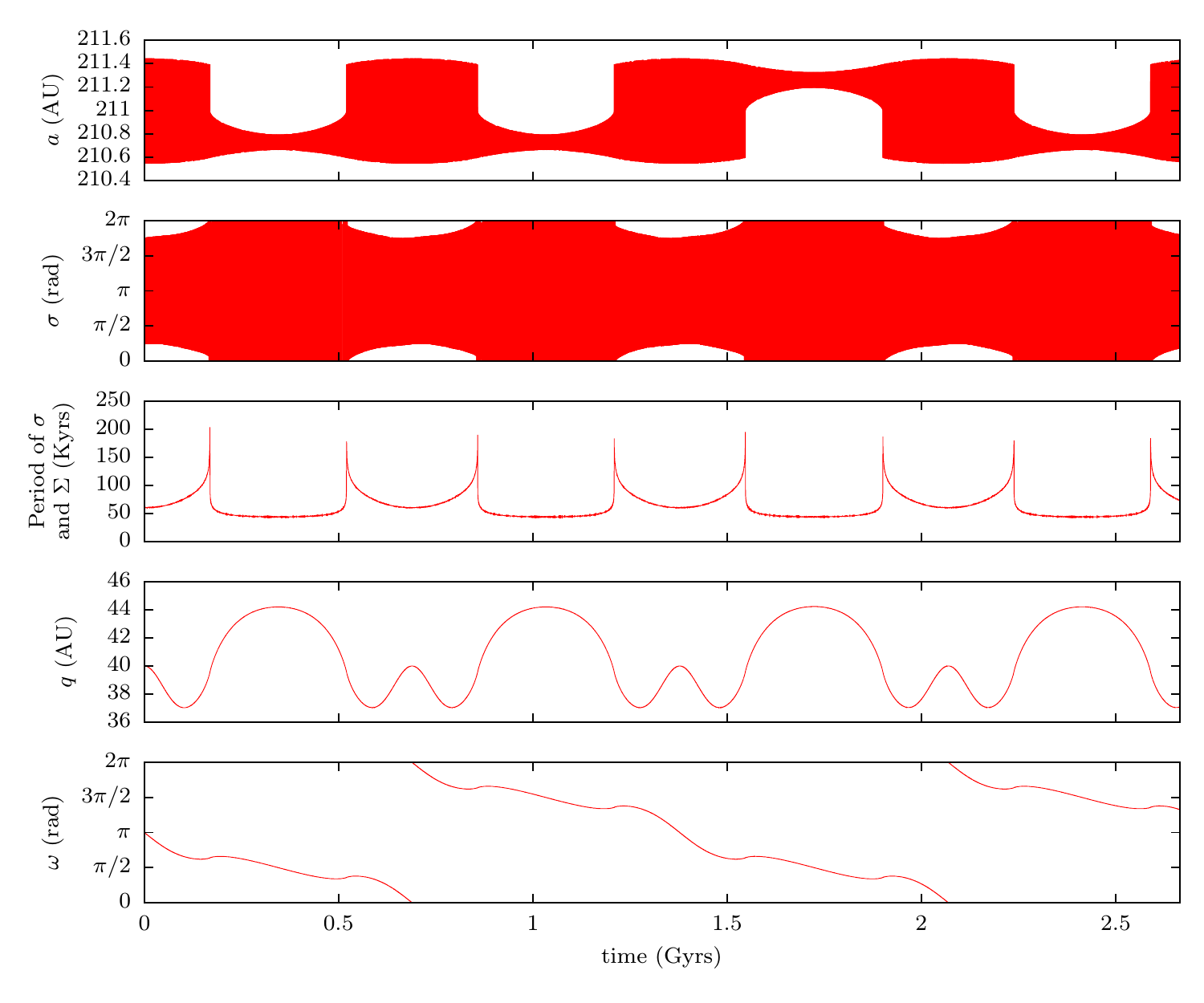}
      \caption{Same as Fig.~\ref{fig:ssCross1-details} but for the trajectory of Fig.~\ref{fig:ssCross2}. The first two and the last circulation phases are toward the right (see $a$ and $\sigma$), whereas the third one is toward the left.}
      \label{fig:ssCross2-details}
   \end{figure}

   Figures~\ref{fig:ssCross1} and \ref{fig:ssCross2} show two examples of such segmented trajectories. Since the diffusion of $J$ is extremely small, we considered only two secular models (one for oscillation, one for circulation) but remember that $J$ is actually not exactly retrieved after each circulation phase. Note that it would be erroneous to superimpose the left and right graphs, because the transition from the oscillation value of $J$ to the circulation one is specific to the red trajectory. On Fig.~\ref{fig:ssCross1}, the circulation phase is rather short and we can easily guess by symmetry the approximative trajectory followed by the particle between the white and black points. This is much less obvious on Fig.~\ref{fig:ssCross2}, where the circulation phase plays an important role in the dynamics. Details of these two semi-secular integrations are given on Fig.~\ref{fig:ssCross1-details} and~\ref{fig:ssCross2-details}. In particular, note the random occurrence of left and right circulation phases with the corresponding central values for the semi-major axis. The secular dynamics is however very similar in both cases: it depends mostly on the amplitude of $J$ and little on its sign. Hence, the right graphs of Fig.~\ref{fig:ssCross1} and~\ref{fig:ssCross2}, which are plotted for a right circulation, correspond also roughly to the ones obtained for a left circulation. Since $J$ is almost exactly recovered after each circulation phase, these trajectories are pretty periodic on a secular time-scale. The separation of the two time-scales can be appreciate on Fig.~\ref{fig:ssCross1-details} and~\ref{fig:ssCross2-details}: the cycles of $\sigma,\Sigma$ (middle graph) run always much faster than the secular evolution of $q$ and $\omega$, even in the neighbourhood of the separatrix crossings.
   
   \subsection{Double islands and $1\!:\!k$ resonances}\label{subsec:double}
      Let us now finish with the most complex case, that is when there are two resonance islands in the $(\Sigma,\sigma)$ plane. According to~\cite{GALLARDO_2006b}, this always happens for resonances of type $1\!:\!k$ provided that the eccentricity is high enough. Moreover, that specific kind of resonances can also admit horseshoe-type orbits enclosing the two oscillation islands. At this point, we can anticipate a bit and look at Fig.~\ref{fig:ss1N5570} for typical examples of such a case. The computation of a secular Hamiltonian as defined previously requires thus an additional choice: once we have set up our minds for an oscillating $\sigma$, what type of orbit do we choose? Oscillations around the left centre, around the right one, or around both of them? The method described in Sec.~\ref{subsec:secres} is valid for each type of trajectories, but it can become a bit tricky to determine numerically the trajectory enclosing the required area.
      
\begin{figure}
         \centering
         \includegraphics[width=0.8\textwidth]{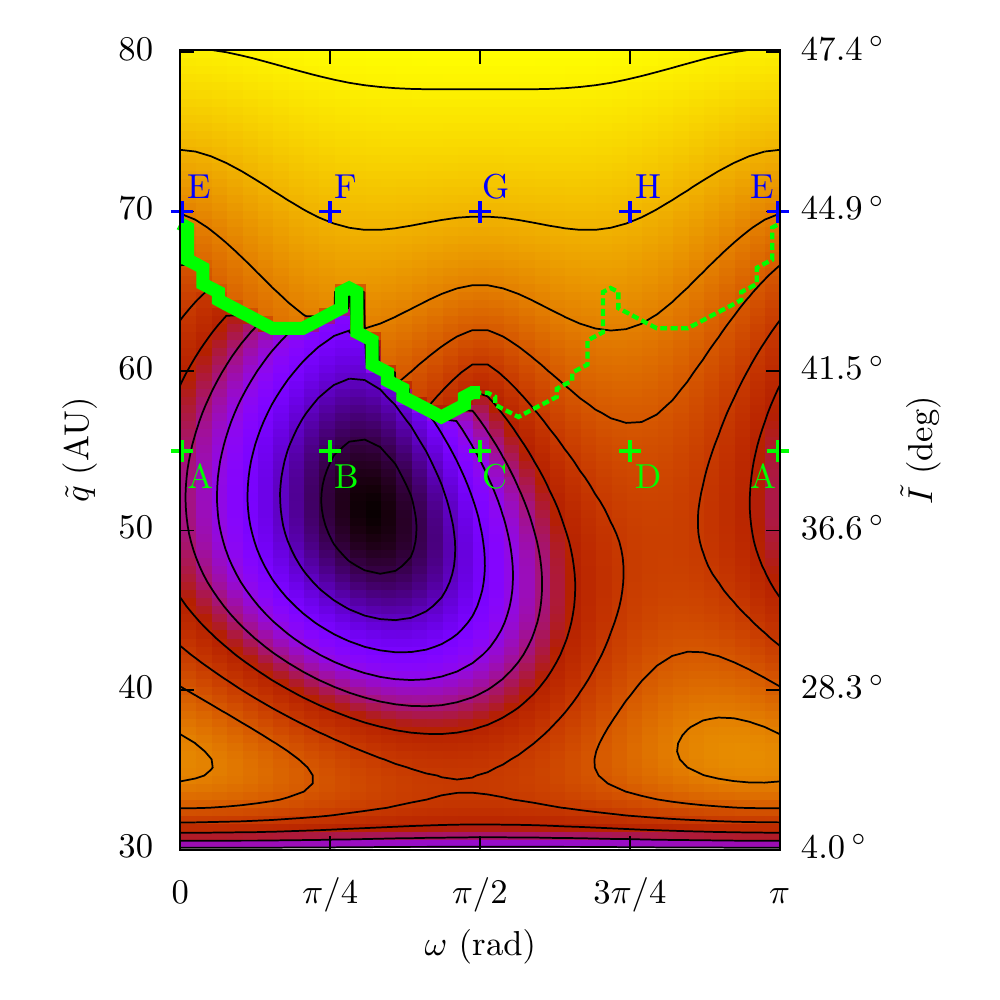}
         \caption{Level curves of the secular Hamiltonian $\mathcal{F}$ for the resonance $1\!:\!11$ with Neptune (reference semi-major axis chosen: $a_0=149.1955$ AU). The parameters are $\eta_0 = 0.6$ and $2\pi J=-3\times 10^{-4}$ AU$^2$rad$^2/$yr. Here, we chose $\sigma$ to oscillate inside the ``right" island (defined from the lower left point of the graph and followed thereafter). Above the green line, only one resonance island remains: for $\omega\in[0,\pi/2]$ it is the remnant of the left island (bold line, discontinuity), but the remnant of the right one for $\omega\in[\pi/2,\pi]$ (thin dashed line, no discontinuity).}
         \label{fig:Fsec_1N}
      \end{figure}

\begin{sidewaysfigure}
         \begin{minipage}[c][12cm][]{\textwidth}\end{minipage}
         \vfill
         \begin{minipage}[c]{\textwidth}
            \centering
            \includegraphics[width=\linewidth]{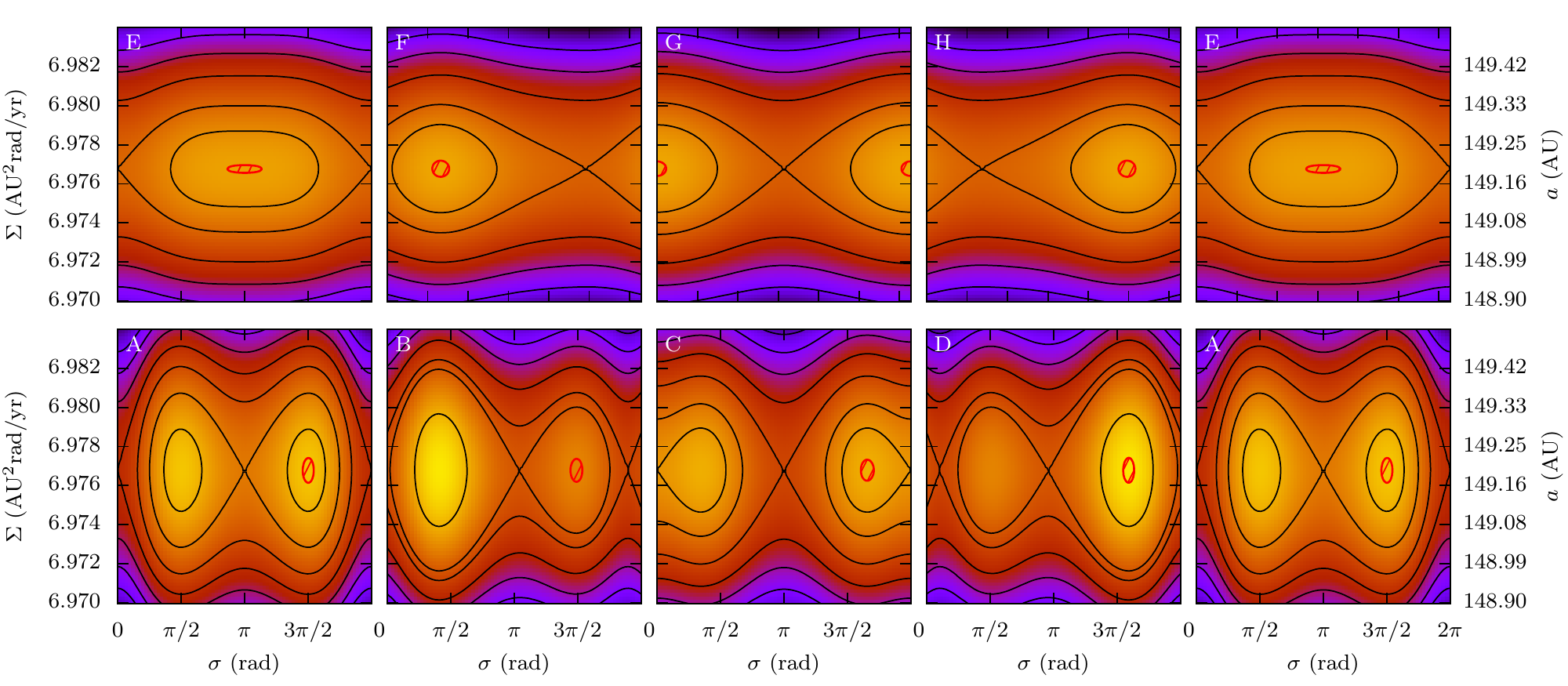}
            \captionof{figure}{Level curves of the semi-secular Hamiltonian $\mathcal{K}$, for $(U,u)$ fixed according to the points A-H of Fig.~\ref{fig:Fsec_1N}. The trajectory enclosing the surface $2\pi J$ is shown in red, and the semi-major axis corresponding to $\Sigma$ is given on the right. Only one resonance island is left for the points E-H because they correspond to a high perihelion distance (see Fig.~\ref{fig:Fsec_1N}).}
            \label{fig:ss1N5570}
         \end{minipage}
      \end{sidewaysfigure}

      Naturally, the geometry of the semi-secular level curves will evolve during the secular evolution of $\omega$ and $\tilde{q}$, and that complicates further the process: the position of the two islands can indeed vary a lot, as well as their sizes. To prevent any accidental jump from an island to another during the numerical computation of the secular levels, we adopted the following strategy:
      \begin{enumerate}
         \item Choose the parameters $\eta_0$ and $J$ (as before), and an oscillation type for $\sigma$ (left, right or horseshoe).
         \item Start the plot from a particular point $(\omega,\tilde{q})$, typically the lower left corner of the graph. We get a first value of $\mathcal{F}$.
         \item Compute the value for the adjacent points \emph{following} the chosen island in the $(\Sigma,\sigma)$ plane. Indeed, since the deformations are continuous, the islands cannot exchange their places between two neighbouring points (assuming a sufficiently fine grid).
         \item Go on with the same procedure for the new points.
      \end{enumerate}
      Naturally, that method is relevant as long as the chosen oscillation type is allowed by the value of $\omega$ and $\tilde{q}$. As a matter of fact, the opening of the horseshoe trajectories can happen quite often during the computation of the secular levels, as well as the disappearance of one of the islands \citep[][defined a critical eccentricity $e_a$ for that]{GALLARDO_2006b}. If that phenomenon happens along a secular trajectory, there \emph{must} be a discontinuity on the plot of $\mathcal{F}$, since the particle is bound to change its type of trajectory. As in Sec.~\ref{subsec:sepcross}, the secular orbits can be defined only by parts, but a secular model is still very informative about the general dynamics. Figure~\ref{fig:Fsec_1N} gives an example of such level curves. As previously, ten points are marked with letters and refer to semi-secular plots (Fig.~\ref{fig:ss1N5570}). The semi-secular Hamiltonian~$\mathcal{K}$ is still symmetric in $\omega$ with respect to $\pi/2$, but this time, the presence of two islands introduces an asymmetry of the secular Hamiltonian~$\mathcal{F}$. We are bound indeed to follow one specific island, as shown on Fig.~\ref{fig:ss1N5570}: the graphs B and D are symmetric but the position of the red trajectory is not. The geometry of the horseshoe-type orbit is even more complicated, since it breaks for $\omega$ slightly greater than $\pi/4$, and reforms with another position near $\omega=\pi/2$ (see the evolution from graph~B to~C).
      
      The disappearance of one island for an increase of the perihelion distance deserves some further comments. For $\omega=0$ and $\omega=\pi/2$, it is obvious that the two islands merge into a single one (compare graphs A-E and C-G). On the contrary, for other values of $\omega$, the $\sigma$-width of the vanishing island decreases until the two saddle points merge into one. The other island remains thus rather unchanged and passes smoothly from a two-island configuration to a single one. This explains the structure of the discontinuity line of Fig.~\ref{fig:Fsec_1N}: depending on the oscillation island occupied by the particle, there can either be a regular transition or a brutal jump to another type of trajectory. In the latter case, the secular model used so far is not relevant any more for that particle, because the definition of $J$ has to be changed. Notice that Figure~\ref{fig:Fsec_1N} has been drawn for a very small value of the area $|2\pi J|$. For bigger values, the transition will simply happen earlier. After the separatrix crossing (see Sec.~\ref{subsec:sepcross}), the particle can either jump around the other centre, follow a horseshoe-type orbit or circulate. As before, the probability of each type of trajectory is random and hard to estimate.
      
\begin{figure}
         \centering
         \includegraphics[width=\textwidth]{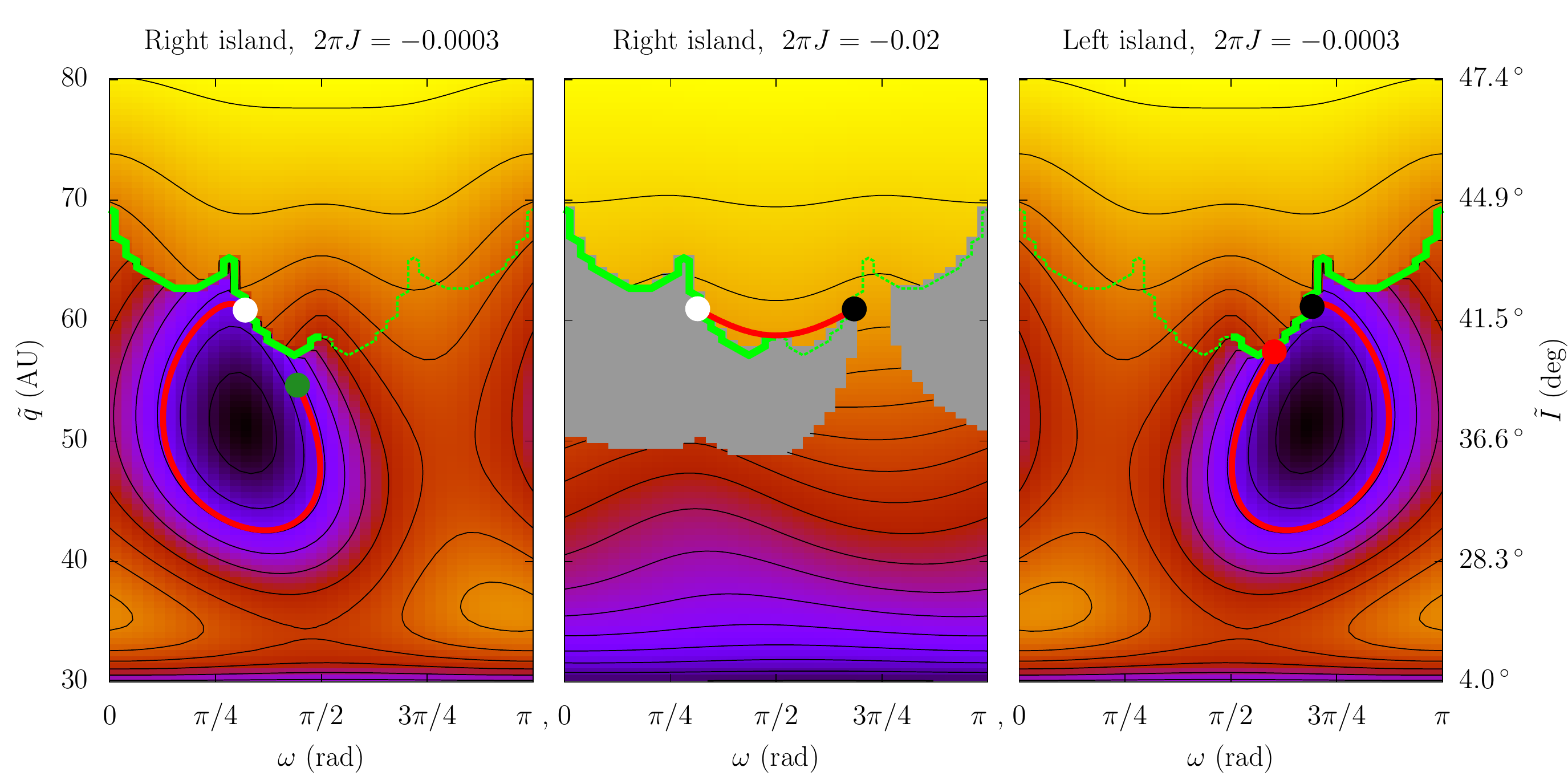}
         \caption{Numerical integration of the two-degree-of-freedom semi-secular system for the resonance $1\!:\!11$ with Neptune. The trajectory is plotted by parts in front of the secular level curves: the common parameter is $\eta_0 = 0.6$ and $J$ is given in AU$^2$rad$^2/$yr above the pictures. It begins with the green spot (left graph), ends with the red (right graph), and follow the colour code in between (white-white and black-black). The middle graph is plotted for oscillations inside the right island, but there is anyway only a single island above the green line. As before, the grey denotes regions where the chosen island is too small to contain the area $|2\pi J|$ required. Since the area is very small for the left and right graphs, the grey band is very thin and hidden under the thick green line.}
         \label{fig:ssEvol1N}
      \end{figure}

\begin{figure}
         \centering
         \includegraphics[width=\textwidth]{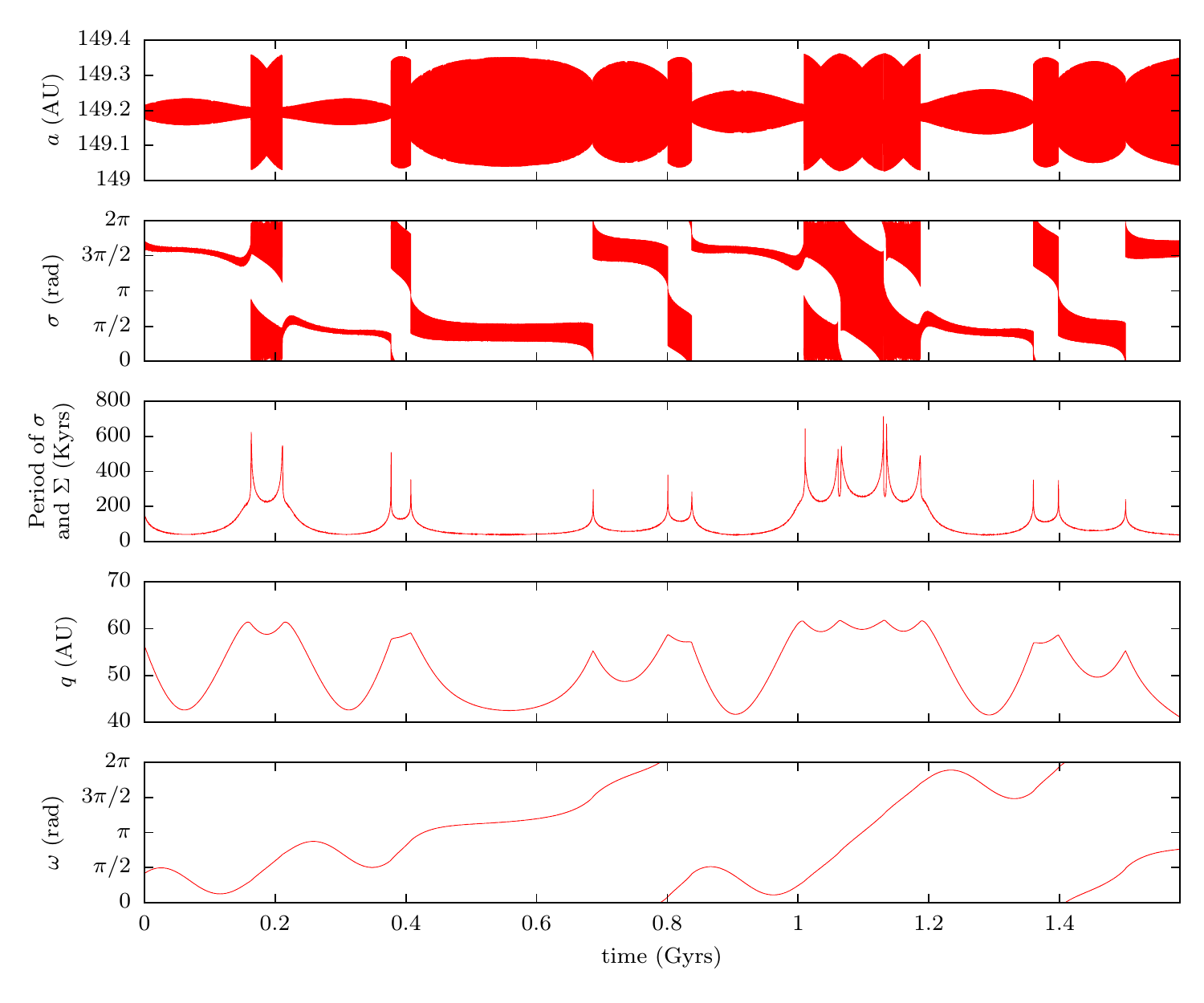}
         \caption{Numerical integration of the two-degree-of-freedom semi-secular system. The semi-major axis is given instead of $\Sigma$ and the perihelion instead of $U$ (see Eq.~\ref{eq:SUV} for the correspondence). The period of $\sigma$ (central or horseshoe oscillation) is given on the middle graph, where the separatrix crossings are obvious (the period tends to infinity). The first three dynamical regimes (see $a$ and $\sigma$ from $t=0$ to $\approx 0.38$ Gyrs) correspond to the trajectory shown on Fig.~\ref{fig:ssEvol1N}.}
         \label{fig:ssEvol1N-det}
      \end{figure}

      As an example, Figure~\ref{fig:ssEvol1N} presents a numerical integration of the semi-secular system for a specific broken level curve of Fig~\ref{fig:Fsec_1N}. The trajectory begins with the green point, where $\sigma$ oscillates inside the right island with a small area. On the white point, the right island vanishes, forcing the particle to follow another type of trajectory. For that particular example, it begins to oscillate in the remaining island with a large area (middle graph). The particle crosses the discontinuity at $\omega=\pi/2$ on a safe horseshoe-type orbit, but hits the growing left island at the black point\footnote{Note that such a symmetrical trajectory was improbable since at the black point the right island is much larger than the left one. A careful analysis of that orbit shows that $\sigma$ is temporarily trapped around the saddle point and then swallowed by the growing left island. However, other integrations of the osculating and semi-secular systems show various possible behaviours, including further stay in the right island with a large area, or temporary maintenance of a grazing horseshoe-type orbit (see for instance Fig.\ref{fig:ssEvol1N-det} just after $1$ Gyr: the double peaks in the period are separatrix approaches without crossing).}. For reasons of symmetry, the adopted area inside the left island is very close to the previous one in the right island. After the red point, the particle goes on switching type of trajectory: see Fig.~\ref{fig:ssEvol1N-det} for the evolution on a wider time-scale. Actually, that kind of behaviour can persist for billion years, as long as the particle does not reach a Neptune-crossing secular trajectory.
      
      Please see~\cite{GALLARDO_2006a} for other examples: its Figure~12 presents a very similar case (same resonance and nearby values of the parameters). Its Figure~13, on the contrary, shows a steadier evolution without separatrix crossing (the particle is locked indefinitely in the ``right" island).
      
      These illustrations show how complicated can be the dynamics inside a $1\!:\!k$ resonance. A secular model can seem rather cumbersome and ineffective for such ``integrable by parts" trajectories, which are chaotic by essence: in fact, it is more designed for general studies about the dynamics than for following a particular realization of it. Note that the chaos invoked here is due to the complex geometries of the $1\!:\!k$ resonances, contrary to a diffusion of the adiabatic invariant as in the case of \cite{WISDOM_1985} discussed above. As seen on Sec.~\ref{subsec:sepcross}, the two time-scales are so well separated that the breaking of the adiabatic hypothesis at separatrix crossings can be considered as instantaneous. Hence, the uncertainty concerns almost solely the new type of trajectory adopted, rather than the new value of~$J$.
      
\section{Discussion and conclusion}
   We presented new tools and results concerning the dynamics of transneptunian objects. When the short term behaviour of the particle is integrable (or quasi-integrable), its long term evolution can be efficiently described by a semi-analytic secular model with one degree of freedom and two parameters. In particular, it proves to be particularly suitable to detect large perihelion excursions.
   
   Such a model is easily obtained when there is no mean-motion resonance in the system \citep{THOMAS-MORBIDELLI_1996}: the two parameters are then $a$ and $C_K=(1-e^2)\cos^2\!I$. The specificity of the region under study allows to derive also an analytical asymptotic model, similar to Kozai's one. Contrary to previous papers, we used these secular representations to get general and quantitative results. For a quasi-integrable non-resonant dynamics, we showed that the maximum perihelion excursion possible is $16.4$ AU, attainable on a Giga-year time-scale for high semi-major axis and a very specific inclination (near $63\degree$ or $114\degree$). A small body beginning with a perihelion near the orbit of Neptune and in the required range of inclination can thus reach rather high perihelion distances from the planets, especially if it has undergone a prior diffusive process. That mechanism, though, cannot explain very large perihelion distances as the ones revealed by massive numerical integrations \citep[][]{GOMES-etal_2005}, or the ones of Sedna and 2012VP$_{113}$ (if we assume for them an initial orbit near the planetary region).
   
   When there is a mean-motion resonance between the body and one of the planets, the adiabatic invariant theory and the coordinate change of~\cite{HENRARD_1990} permitted to construct a new accurate secular model, similar to the non-resonant one. The two fixed parameters are then $V$ (a surrogate of $C_K$) and the secular area $2\pi J$ enclosed by the librating resonant canonical coordinates. The only obstruction to a fully integrable representation comes then from a possible extreme narrowing of the resonance island, which can make circulate the resonant angle (separatrix crossing). In such case, a secular representation is possible only by parts, each of them with a different parameter $J$. Such transitions can happen frequently for the resonances of type $1\!:\!k$, even for enclosed areas equal to zero (disappearance of the resonance island). For a specific trajectory, these repeated changes of behaviour are an evident source of long term chaos and make somehow questionable the use of a secular model. It remains though very effective as a general tool, to locate the secular equilibrium points and distinguish in a glance the regular trajectories from the ``segmented" ones.
   
   With our resonant model, it was straightforward to bring out some trajectories with very large perihelion variations (for instance from $30$ to $80$ AU). Such extreme values were usually considered too high to be reached by the means of perturbations from the known planets \citep[see for example][]{BROWN-etal_2004}. Moreover, that kind of trajectory is not restricted to high-inclination regimes as in the non-resonant case. Further applications, using parameters of known objects and exploring the parameter space, are kept for future papers. It would not be surprising, for instance, to detect some secular equilibrium points common to numerous different resonances and corresponding to accumulation values of $\omega$ in the observed distribution of transneptunian objects. That kind of secular theories could also be applied to extended models of the Solar System including possible distant planets still undiscovered \citep[see for instance][]{BATYGIN-BROWN_2016}. The comparison of the dynamical paths allowed by the different models could thus precise if such new features are indeed necessary to explain the observed distribution of the distant transneptunian objects.

\begin{acknowledgements}
   We thank Giovanni~F. Gronchi and Andrea Milani for their precious advice: sometimes, even a few words can be of great help. We thank also two anonymous referees who helped us to improve the paper.
\end{acknowledgements}

\bibliographystyle{aps-nameyear} % American Physical Society (APS) style, author-year citations
\bibliography{CM_2015}

\normalsize
\appendix
\section{Expansion of the secular non-resonant Hamiltonian}
   We present here some hints about the construction of the secular analytical non-resonant Hamiltonian function. With the planetary model chosen (Eq.~\ref{eq:pmod}), the angle $\psi_i$ is simply defined by:
   \begin{equation}
      \cos\psi_i =\alpha\cos\lambda_i + \beta\sin\lambda_i
   \end{equation}
   where:
   \begin{equation}
      \left\{
      \begin{aligned}
        \alpha &= \cos(\omega+\nu)\cos\Omega-\sin(\omega+\nu)\sin\Omega\cos I\\
        \beta  &= \cos(\omega+\nu)\sin\Omega+\sin(\omega+\nu)\cos\Omega\cos I
      \end{aligned}
      \right.
   \end{equation}
   Then, the multiple average of $\varepsilon\mathcal{H}_1$ is computed in two steps, beginning with the integration over $\lambda_1,\lambda_2...\lambda_N$. As the indirect part vanishes, we have, using the Legendre development~\eqref{eq:devPL}:
   \begin{equation}\label{eq:intlbd}
      \frac{1}{(2\pi)^N}\int_{0}^{2\pi}\!\!\!\int_{0}^{2\pi}\!\!\!\!\!...\int_{0}^{2\pi}\varepsilon\mathcal{H}_1\ \ \mathrm{d}\lambda_1\mathrm{d}\lambda_2...\mathrm{d}\lambda_N = -\frac{1}{r}\,\sum_{n=0}^{+\infty}\left(\sum_{i=1}^{N}\mu_i\left(\frac{a_i}{r}\right)^{2n}\right)P_{2n}(\chi)
   \end{equation}
   where $\chi$ is defined as:
   \begin{equation}
      \left\{
      \begin{aligned}
         \chi^0 &= 1 \\
         \chi^{2k} &= \frac{1\times 3\times 5\times...\times(2k-1)}{2\times 4\times 6\times...\times 2k}(\alpha^2+\beta^2)^{k}\text{\ \ \ ,\ \ \ } k = 1,2,3...
      \end{aligned}
      \right.
   \end{equation}
   One can see that the odd terms have disappeared. The average over $l$ is less straightforward, since each polynomial $2n$ of \eqref{eq:intlbd} requires the computation of $n+1$ integrals of the form:
   \begin{equation}
      \frac{1}{2\pi}\int_{0}^{2\pi}\frac{(\alpha^2+\beta^2)^k}{r^{2n+1}}\,\mathrm{d}l
      = \frac{1}{a^2\sqrt{1-e^2}}\ \frac{1}{2\pi}\int_{0}^{2\pi}\frac{(\alpha^2+\beta^2)^k}{r^{2n-1}}\,\mathrm{d}\nu
      \text{\ \ \ ,\ \ \ }k = 0,1,2...n
   \end{equation}
   where $\alpha$, $\beta$ and $r$ are functions of the true anomaly $\nu$. The general form of the result is presented in equations~\eqref{eq:eF1} and~\eqref{eq:Bn}, and the first terms are the followings:
   \begin{equation*}
      \begin{array}{|l|}
      \hline
         \multicolumn{1}{|c|}{n = 1} \\
         \hline \hline
         \multicolumn{1}{|c|}{\alpha_1 = 1/8} \\
         \hline \hline
         P_1^0(x) = 1 \\
         Q_1^0(x) = -1+3\,x^2 \\
      \hline
      \end{array}
      \hspace{0.5cm}
      \begin{array}{|l|}
      \hline
         \multicolumn{1}{|c|}{n = 2} \\
         \hline \hline
         \multicolumn{1}{|c|}{\alpha_2 = 9/1024} \\
         \hline \hline
         P_2^0(x) = 2+3\,x^2 \\
         Q_2^0(x) = 3-30\,x^2+35\,x^4 \\
         \hline
         P_2^1(x) = 10 \\
         Q_2^1(x) = -1+7\,x^2 \\
      \hline
      \end{array}
      \hspace{0.5cm}
      \begin{array}{|l|}
      \hline
         \multicolumn{1}{|c|}{n = 3} \\
         \hline \hline
         \multicolumn{1}{|c|}{\alpha_3 = 25/65536} \\
         \hline \hline
         P_3^0(x) = 2\,(8+40\,x^2+15\,x^4) \\
         Q_3^0(x) = -5+105\,x^2-315\,x^4+231\,x^6 \\
         \hline
         P_3^1(x) = 210\,(2+x^2) \\
         Q_3^1(x) = 1-18\,x^2+33\,x^4 \\
         \hline
         P_3^2(x) = 63 \\
         Q_3^2(x) = -1+11\,x^2 \\
         \hline
      \end{array}
   \end{equation*}
   \vspace{0.5cm}
   \begin{equation*}
      \begin{array}{|l|}
      \hline
         \multicolumn{1}{|c|}{n = 4} \\
         \hline \hline
         \multicolumn{1}{|c|}{\alpha_4 = 245/33554432} \\
         \hline \hline
         P_4^0(x) = 5\,(16+168\,x^2+210\,x^4+35\,x^6) \\
         Q_4^0(x) = 35-1260\,x^2+6930\,x^4-12012\,x^6+6435\,x^8 \\
         \hline
         P_4^1(x) = 630\,(48+80\,x^2+15\,x^4) \\
         Q_4^1(x) = -1+33\,x^2-143\,x^4+143\,x^6 \\
         \hline
         P_4^2(x) = 1386\,(10+3\,x^2) \\
         Q_4^2(x) = 1-26\,x^2+65\,x^4 \\
         \hline
         P_4^3(x) = 858 \\
         Q_4^3(x) = -1+15\,x^2 \\
         \hline
      \end{array}
   \end{equation*}
   \vspace{0.5cm}
   \begin{equation*}
      \begin{array}{|l|}
      \hline
         \multicolumn{1}{|c|}{n = 5} \\
         \hline \hline
         \multicolumn{1}{|c|}{\alpha_5 = 567/4294967296} \\
         \hline \hline
         P_5^0(x) = 14\,(128+2304\,x^2+6048\,x^4+3360\,x^6+315\,x^8) \\
         Q_5^0(x) = -63+3465\,x^2-30030\,x^4+90090\,x^6-109395\,x^8+46189\,x^{10} \\
         \hline
         P_5^1(x) = 9240\,(32+112\,x^2+70\,x^4+7\,x^6) \\
         Q_5^1(x) = 7-364\,x^2+2730\,x^4-6188\,x^6+4199\,x^8 \\
         \hline
         P_5^2(x) = 240240\,(8+8\,x^2+x^4) \\
         Q_5^2(x) = -1+45\,x^2-255\,x^4+323\,x^6 \\
         \hline
         P_5^3(x) = 8580\,(14+3\,x^2) \\
         Q_5^3(x) = 3-102\,x^2+323\,x^4 \\
         \hline
         P_5^4(x) = 12155 \\
         Q_5^4(x) = -1+19\,x^2 \\
         \hline
      \end{array}
   \end{equation*}
   \vspace{0.5cm}
   \begin{equation*}
      \begin{array}{|l|}
      \hline
         \multicolumn{1}{|c|}{n = 6} \\
         \hline \hline
         \multicolumn{1}{|c|}{\alpha_6 = 7623/549755813888} \\
         \hline
         P_6^0(x) = 14\,(256+7040\,x^2+31690\,x^4+36960\,x^6+11550\,x^8+693\,x^{10}) \\
         Q_6^0(x) = 231-18018\,x^2+225225\,x^4-1021020\,x^6+2078505\,x^8-1939938\,x^{10}+676039\,x^{12} \\
         \hline
         P_6^1(x) = 60060\,(128+768\,x^2+1008\,x^4+336\,x^2+21\,x^8) \\
         Q_6^1(x) = -3+225\,x^2-2550\,x^4+9690\,x^6-14535\,x^8+7429\,x^{10} \\
         \hline
         P_6^2(x) = 90090\,(80+168\,x^2+70\,x^4+5\,x^6) \\
         Q_6^2(x) = 5-340\,x^2+3230\,x^4-9044\,x^6+7429\,x^8 \\
         \hline
         P_6^3(x) = 12155\,(224+160\,x^2+15\,x^4) \\
         Q_6^3(x) = -5+285\,x^2-1995\,x^4+3059\,x^6 \\
         \hline
         P_6^4(x) = 230945\,(6+x^2) \\
         Q_6^4(x) = 1-42\,x^2+161\,x^4 \\
         \hline
         P_6^5(x) = 29393 \\
         Q_6^5(x) = -1+23\,x^2 \\
         \hline
      \end{array}
   \end{equation*}
\end{document}